\pdfoutput=1
\documentclass[aps,pre,twocolumn,superscriptaddress,longbibliography]{revtex4-1}
\usepackage[english]{babel}
\usepackage{amsmath, amssymb, mathtools, bbm}
\usepackage{graphicx, tikz, pifont}
\newcommand{\cmark}{\ding{51}}%
\newcommand{\xmark}{\ding{55}}%
\usepackage[babel]{csquotes}
\usepackage{natbib}
\usepackage{url}
\usepackage[colorlinks]{hyperref}

\begin{document}

\title{A framework of nonequilibrium statistical mechanics. II. Coarse-Graining}

\author{Alberto Montefusco}
\affiliation{ETH Z\"{u}rich, Department of Materials, Polymer Physics, CH-8093 Z\"urich, Switzerland}
\author{Mark A. Peletier}
\affiliation{Technische Universiteit Eindhoven, Centre for Analysis, Scientific Computing and Applications, and Institute for Complex Molecular Systems (ICMS), 5600 MB Eindhoven, The Netherlands}
\author{Hans Christian \"{O}ttinger}
\affiliation{ETH Z\"{u}rich, Department of Materials, Polymer Physics, CH-8093 Z\"urich, Switzerland}
\date{\today}

\begin{abstract}
 For a given thermodynamic system, and a given choice of coarse-grained state variables, the knowledge of a force-flux constitutive law is the basis for any nonequilibrium modeling. In the first paper of this series we established how, by a generalization of the classical fluctuation-dissipation theorem (FDT), the structure of a constitutive law is directly related to the distribution of the fluctuations of the state variables. When these fluctuations can be expressed in terms of diffusion processes, one may use Green-Kubo-type coarse-graining schemes to find the constitutive laws. In this paper we propose a coarse-graining method that is valid when the fluctuations are described by means of general Markov processes, which include diffusions as a special case. We prove the success of the method by numerically computing the constitutive law for a simple chemical reaction $A \rightleftarrows B$. Furthermore, we show that one cannot find a consistent constitutive law by any Green-Kubo-like scheme.
\end{abstract}

\maketitle

\section{Introduction}
 In nonequilibrium thermodynamics, fluctuations and dissipation are two sides of the same coin: the dissipative structure (i.e., the force-flux constitutive law) of the phenomenological equations is in a one-to-one correspondence with the distribution of their fluctuations, which represent, as an idealization, the effect of more microscopic, neglected, fast degrees of freedom. This is the essence of the `fluctuation-dissipation theorem of the second kind', according to the classification designed by Kubo and co-authors \cite[Sec.~1.6]{KTH85}.
 
 The fluctuation-dissipation theorem of the second kind (henceforth shortened as FDT) is the main subject of this series of papers. It has many uses, and  we discuss two of these here. First, if we know the dissipative structure of the phenomenological equations, we can construct fluctuations that are compatible with this dissipative structure; this operation is performed, for instance, to augment the equations of hydrodynamics with fluctuating fluxes \cite[§~88]{LL13}, \cite{pE98,BGDB14}, and we have called this an \emph{enhancement} in the first paper of this series (which we refer to as (I)). Secondly, in the opposite direction, if the dissipative structure of the phenomenological equations is not yet known, then we can determine it by analyzing the fluctuations that result from the \emph{coarse-graining} of a more microscopic model: this is a major task of nonequilibrium statistical mechanics \cite[{Chapter~6}]{hcO05}, \cite{hcO07}. This second use of the fluctuation-dissipation is the topic of the current paper.
 
 In~(I) we observed how Kubo's classical formulation of the FDT is limited to fluctuations described by diffusion processes and, inspired by \cite{MPR14,MielkePeletierRenger16}, postulated an extension of the FDT to general Markov processes, and corresponding extended theories of enhancements and dynamic coarse-graining. In this second paper we focus on coarse-graining where, by this expression, we mean (i)~the identification of good macroscopic variables for a given problem, and (ii)~the determination of the emerging dissipative structure at a chosen, more macroscopic level of description, which is our main focus. Step~(i), namely the identification of good macroscopic variables \cite{VTP16}, is usually problem-dependent and requires a great deal of experience: it can hardly be framed in a systematic method, as we do here for step~(ii).
 
 In the setting of diffusive fluctuations, the method of Green-Kubo relations, based on the classical FDT, has been established as a powerful coarse-graining tool~\cite{hcO07}, in that it allows us to compute a dissipative structure, expressed in terms of a friction matrix, by relatively short simulations and without the need of imposing a different external force for each irreversible process. The hydrodynamics of Newtonian fluids and Fourier's heat conduction are the prime examples. The method has been used both in the linear-response regime \cite[Section~4.4]{EM07} and in fully nonequilibrium situations \cite{KO04,IOK09,IMO10}. The diffusive nature of the fluctuations is a feature of state variables that are the sum of short-time correlated interactions of many microscopic particles~\cite{OM53} and evolve continuously in time by infinitesimally small movements in state space.
 
 One of the main messages of (I) was that the framework of Green-Kubo relations is not general enough. Indeed, it does not include macroscopic variables characterized by rare events, for which, in contrast to the diffusive case, it is always possible to find a time scale at which the dynamics appears as constituted of rare and sudden jumps at discrete instants of time. The chief example is a chemical reaction, for which the fluctuating dynamics resembles more a jump process than a diffusion. For such systems, the picture identified by Eyring~\cite{hE35} and Kramers~\cite{haK40} and the corresponding rate formula gave birth to the field of rare-event estimation and simulation \cite{jaB04,eVE06,HBSBS14,FVEW15,BA16}. However, none of the methods in this field can resolve the dissipative structure of the macroscopic phenomenological equations. Trying to apply the Green-Kubo method to such systems produces incorrect results, as we show in Sec.~\ref{Green-Kubo}. Indeed, the theory presented in (I) shows how, for such systems, the correct dissipative structures may be expressed in terms of dissipation potentials instead of friction matrices.
 
 The purpose of this paper is to construct an extended theory of coarse-graining based on the generalized FDT that we proposed in (I). The theory allows us to resolve dissipative structures of the phenomenological equations associated with (diffusive and non-diffusive) Markov processes and thus unifies the two pictures, Green-Kubo relations and Kramers-like rate formulas. We show how this works in practice by numerically computing the dissipation potential for the rate equation of the elementary chemical reaction $A \leftrightarrows B$, which arises from the coarse-graining of a dynamics in an double-well energy landscape (the Kramers escape problem).
 
 One reason for determining the dissipative structure of a phenomenological equation has already been mentioned: it allows us to identify the type of noise that gives a consistent stochastic \emph{enhancement} of the phenomenological equations, as we described in (I). The ultimate decision about the proper dissipative structure and the corresponding noise enhancement can only be reached by analyzing the true fast microscopic dynamics, that is, by statistical mechanics. However, if we have some intuition about the qualitative features of the noise, we can fill in the quantitative details of the idealized Markovian noise from the macroscopic dissipative properties. Conversely, the dissipative structure guides us in making statistical mechanics efficient for extracting all the dissipative properties from microscopic simulations. As the result of a consolidation process we arrive at a consistent multiscale understanding of a given system of interest.
  
 The paper is structured as follows. After recapitulating, in Sec.~\ref{sec:Gaussian}, the classical Green-Kubo method for diffusive fluctuations, we propose, in Sec.~\ref{sec:generalized}, the natural extension to Markov processes on the basis of the generalized FDT. In Sec.~\ref{sec:Kramers} we use the extended method to numerically compute the dissipation potential associated to the chemical reaction $A \leftrightarrows B$, and we present our conclusions in Sec.~\ref{sec:conclusions}.
 
\section{Coarse-graining associated with diffusion processes}\label{sec:Gaussian}
 \subsection{Inference for diffusion processes}
  Let us consider the stochastic differential equation (SDE)
  \begin{equation}\label{diffusion}
   d X_t = A(X_t) \, d t + B(X_t) \diamond d W_t \, ,
  \end{equation}
  with unknown drift~$A(x)$, noise intensity~$B(x)$ with \emph{Klimontovich} interpretation (cf.~the generator~\eqref{generatorDiffusion}) \cite{ylK90}, and $x \in \mathbb{R}^d$. The estimation of $A(x)$ and $B(x)$, or the corresponding diffusion matrix $D(x) \coloneqq B(x) B(x)^T$, is the general problem of inference for diffusion processes, which in the statistics literature is studied for instance in \cite{smI08,cF13}.
 
  For our illustrative purposes, working in terms of the infinitesimal generator of the process \eqref{diffusion},
  \begin{equation}\label{generatorDiffusion}
   (\mathcal{Q} f)(x) = \frac{\partial f(x)}{\partial x} \cdot A(x) + \dfrac{1}{2} \frac{\partial}{\partial x} \cdot \left[ D(x) \frac{\partial f(x)}{\partial x} \right] ,
  \end{equation}
  is particularly transparent. Indeed, applying the generator to the test functions $f_i(x) = x_i$ and $f_{ij}(x) = x_i x_j$, we see that the diffusion matrix may be computed by the formula
  \begin{align}\label{GKgenerator}
   D_{i j}(x) &= (\mathcal{Q} f_{i j})(x) - x_i \, (\mathcal{Q} f_j)(x) - x_j \, (\mathcal{Q} f_i)(x) \\
   &= \lim\limits_{\tau \to 0} \dfrac{1}{\tau} \mathbb{E}\!\left[ \left( (X_\tau)_i - x_i \right) \left( (X_\tau)_j - x_j \right) \Big\vert X_0 = x \right] , \nonumber
  \end{align}
  or, in matrix notation,
  \begin{equation}\label{GK1}
   D(x) = \lim\limits_{\tau \to 0} \dfrac{1}{\tau} \mathbb{E}\!\left[ \left( X_\tau - x \right) \left( X_\tau - x \right)^T \Big\vert X_0 = x \right] .
  \end{equation}
  As a consequence, the drift is computed as \footnote{The presence of the divergence of $D$ is due to the Klimontovich interpretation for the noise.}
  \begin{equation}
   A_i(x) = (\mathcal{Q} f_i)(x) - \dfrac{1}{2} \sum\limits_{j=1}^d \partial_j D_{i j}(x) \, ,
  \end{equation}
  or
  \begin{equation}
   A(x) = \lim\limits_{\tau \to 0} \dfrac{1}{\tau} \left( \mathbb{E}\!\left[X_\tau \big\vert X_0 = x\right] - x \right) - \dfrac{1}{2} \frac{\partial}{\partial x} \cdot D(x) \, .
  \end{equation}
  
  All formulas inspired by Eq.~\eqref{GK1} are referred to as \emph{Green-Kubo formulas} and assume different forms depending on the application \cite[Sec.~8.4]{hcO05}. We consider them as a way to infer the diffusion matrix from a sample of the time correlations of the second moments (cf.~\cite[Sec.~4.2.3]{smI08}). Although we have considered only processes in $\mathbb{R}^d$, the same conclusions can be drawn, at least formally, for infinite-dimensional systems; see \cite{daD93} for a theory of measure-valued diffusion processes, and \cite{EDZR18} for an application to a field theory.
  
 \subsection{Coarse-graining via the Green-Kubo formula}\label{sec:GK}
Let a physical system be described by the variables $y = \left(y^1, y^2, \ldots, y^n\right)$, with a large number~$n$ of degrees of freedom, and suppose that its dynamics is characterized by a `slow' and a `fast' time scale. The first goal of dynamic coarse-graining, in the sense used in this paper, is to identify a set of more macroscopic state variables~$X(y) \in \mathbb{R}^d$ that resolve the slow dynamics, meaning that, in the limit of many degrees of freedom, they evolve only `slowly'. In addition, we suppose that the dynamics of the macroscopic variables, for large but finite~$n$, can be well approximated by a stochastic process, which we also call~$X$ for simplicity: the most probable realizations of the stochastic process are regarded as paths of the slow dynamics, and the noise represents the idealized effect, at the macroscopic level, of the more microscopic, fast degrees of freedom. The identification of the coarse-grained variables (also called ``collective variables'') is often a very challenging step~\cite{VTP16}.
  
In many cases the macroscopic variables $X$ can be assumed to be the sum of short-time correlated interactions of many microscopic particles~\cite{OM53}. The short-time correlation implies that $X$ can be assumed to be Markovian, in the presence of an infinitely large separation of time scales; in practice one has to deal with finite separations, which lead to memory effects. When the interactions are frequent and small, such that the process~$X$ can be considered to have continuous sample paths, then the process~$X$ is a \emph{diffusion} and can be described by an SDE~\cite[Sec.~2.5]{Pavliotis14}. In this case the noise is Gaussian.

  \smallskip
  
  We saw in (I) how the classical FDT suggests that the diffusion process~$X_t$ solves the SDE
  \begin{equation}\label{diffusion-FDT}
   d X_t = M(X_t) \frac{\partial S(X_t)}{\partial x} \, d t + B(X_t) \diamond d W_t \, ,
  \end{equation}
  with $2 k_B M(x) = D(x)$; the operator~$M(x)$ is called a \emph{friction matrix} \cite{GO97,OG97}, and the most probable path solves the gradient-flow equation
  \begin{equation}
   \dfrac{d x_t}{d t} = M(x_t) \frac{\partial S(x_t)}{\partial x} \, .
  \end{equation}
  In contrast to (I), here we are assuming only dissipative components in the slow dynamics. This assumption is related to a condition of detailed balance \cite[Sec.~ II.5]{tmL05} of the stochastic process~$X_t$ with respect to the distribution $e^{S(x)/k_B}$; see \cite[Sec.~6.3.5]{cG09} and \cite{MPR14}.
  
  The second goal of dynamic coarse-graining, for such systems, is to infer the dissipative structure, encoded by $S(x)$ and $M(x)$, by simulations. In this paper we always assume that the static simulations for the estimation of $S(x)$ have already been performed, and the dynamic simulations for evaluating the friction matrix $M(x)$ are based on the Green-Kubo formula \eqref{GK1}.
  
  Two remarks are in order. On the one hand, in the Green-Kubo formula, the limit $\tau \to 0$ presupposes a continuous-time sample, while in concrete applications we always consider discrete-time observations. On the other hand, the limit $\tau \to 0$ is not even desirable, since the diffusion process~$X_t$ is only an approximation of the true (in general non-Markovian) macroscopic dynamics, and this approximation is usually valid in a range of time scales that are large with respect to the microscopic, fast time scale~$t_2$. The limit $\tau \to 0$ is thus replaced by $\tau$ being in the range
  \begin{equation*}
  t_2 \ll \tau \ll t_1 \, ,
  \end{equation*}
  where $t_1$ is the macroscopic, slow time scale. In other words, $\tau$ has to be \emph{macroscopically small}, but \emph{microscopically large}. Hence, the friction matrix may be estimated as
  \begin{equation}\label{GK}
   M(x) = \dfrac{1}{2 k_B \tau} \mathbb{E}\!\left[ \left( X_\tau - x \right) \left( X_\tau - x \right)^T \Big\vert X_0 = x \right] .
  \end{equation}
  
  By combining coarse-graining with the fluctuation-dissipation theorem, as described above, many systems have successfully been studied. On the side of theory, we refer to \cite[Sec.~4.2]{KTH85}, \cite[Sec.~2.7]{AT17}, \cite[Sec.~8.4]{hcO05}, \cite{hcO07}, \cite[Chapter~4]{EM07}; for numerical results, see e.g.\ \cite{MG97,GZZ03,EDZR18}. The power of the method stems from two important features: (i) short simulations, with respect to the slow time scale~$t_1$, are sufficient to sample the matrix $M(x)$; (ii) a single numerical experiment provides us with both the evolution equations and their dissipative structure, without the need of resorting to a different experiment for each irreversible process.
  
  \smallskip
    
  However, as we have remarked in (I), some systems have fluctuations that are distinctly non-Gaussian in nature, suggesting that the method just described may not give correct results. Our main example in Section~\ref{sec:Kramers} below is of this type. In that section we apply and implement the generalized FDT of this paper to this example, but we also investigate to which extent the  application of Gaussian-based methods would give incorrect results (see Section~\ref{Green-Kubo}).
  
  In Section~\ref{Green-Kubo} we study three Gaussian-based methods: `Green-Kubo', based on the scheme of this section, `chemical Langevin equation', and `log-mean equation'. Each of them is defined by a different choice of the pair $\left(S(x), M(x)\right)$. It is possible to sample this pair correctly for the Green-Kubo method, but with the result of a wrong macroscopic phenomenological equation; for the other two methods the macroscopic evolution equation is correct, but a simulation would fail in sampling either $S(x)$ or $M(x)$.
  
  Apparently, applying Gaussian-based methods to systems with non-Gaussian fluctuations is like forcing a square peg into a round hole: the results will be suboptimal. In order to deduce the dissipative structure of such non-Gaussian systems, in this series of papers, we generalize the FDT and propose an extended theory of coarse-graining.
  
\section{Coarse-graining associated with Markovian systems}\label{sec:generalized}
 This section addresses the main contribution of this paper: how we can construct a coarse-graining procedure that gives us the dissipative structure of phenomenological equations associated with both Gaussian and non-Gaussian noise. We present the arguments in two subsections. In the first one we briefly define the classical problem of inference for Markov jump processes, and sketch the setup where a dynamics in an energy landscape with metastable states is coarse-grained to a jump process, with unknown, to-be-computed, transition rates; the macroscopic trajectories, namely the solutions of the phenomenological equations, are the most probable paths. In the second subsection we show how the generalized FDT is essential to resolve the dissipative structure of the phenomenological equations.
 
 \subsection{Inference for Markov jump processes}\label{sec:Markov}
  Markov jump processes are completely defined by the transition rates between all of their states, and this information is encoded in the \emph{transition-rate matrix}, fully equivalent to the infinitesimal generator. Very often, a Markov jump process is observed during an experiment or a numerical simulation, and one wishes to determine the transition rates from the observations. This is the task of the statistical inference for Markov jump processes, which was studied for instance in \cite{BS05} and reviewed in \cite{MDJS07}.
  
  In the case of continuous-time observations, it is easy to check \cite{BS05} that the best estimator for an element of the transition-rate matrix is
  \begin{equation*}
   q_{ij} = N_{ij}/R_i \quad (i \neq j) \, ,
  \end{equation*}
  where $N_{ij}$ is the number of transitions from the state $i$ to $j$, and $R_i$ is the time spent in $i$.
  
  The observations, however, are never continuous, but are made at discrete times. The main issue of this situation is the famous \emph{embedding problem} \cite{jfcK62,BS05}: to the same discrete-time Markov chain, the transition rates of which are estimated in practice, there may correspond zero, one, or many continuous-time Markov processes that have the same finite-time transition rates. We will not focus on this issue in this paper, although we believe it to be important for future developments.
  
  \smallskip
  
  The typical physical setup where jump processes arise naturally is the dynamics of many particles in an energy landscape characterized by metastable states. By a \emph{metastable} state one indicates the region of attraction of a local energy minimum such that the time scale for the system to equilibrate is much shorter than the time scale to escape from it \cite{BdH15,DLLN16}. Since the system spends most of its time in these special states, the escape events are `rare', and the study of these phenomena has given rise to the field of \emph{rare-event} estimation and simulation \cite{jaB04,eVE06,HBSBS14,FVEW15,BA16}.
  
%
  The most famous, preliminary toy model in this class of systems was studied by Eyring \cite{hE35} and Kramers \cite{haK40}, who gave the statistical-mechanical derivation of an explicit formula for the transition rates between the two minima of a double-well potential. We will study this system again from the standpoint of our generalized FDT in Sec.~\ref{sec:Kramers}.
  
  In particular, our goal is not to compute the transition rates between metastable states, but to resolve the dissipative structure of the macroscopic phenomenological equations, as we explore in the next subsection.
  
 \subsection{Coarse-graining via the generalized FDT}
  By analogy with the Gaussian picture of Sec.~\ref{sec:GK}, let us consider again a setup described by microscopic variables~$y$, and with a separation of time scales. A set of more macroscopic variables $X(y)$ is introduced to separate the time scales effectively. We now assume that $X$ is not necessarily a diffusion process, but may be represented as a general Markov process, and the most probable paths are the solutions of the deterministic equation
  \begin{equation}\label{deterministic}
   \dfrac{d x_t}{d t} = A(x_t) \, .
  \end{equation}
  We then aim to estimate the structure of this phenomenological equation by analyzing the noise that results, at the macroscopic level, from the neglected degrees of freedom. Among these two steps, identification of the macroscopic variables and estimation of the force-flux constitutive law, which together define what we call a coarse-graining procedure, this paper focuses entirely on the second one.
  
  In contrast to paper (I), here we restrict the form of the phenomenological equations to the generalized gradient flow
  \begin{equation}
   \dfrac{d x_t}{d t} = \left. \frac{\partial \Psi^*(x_t, \xi_t)}{\partial \xi} \right\vert_{\xi_t = \tfrac{\partial S(x_t)}{\partial x}} \, ,
  \end{equation}
  which corresponds to a purely dissipative dynamics governed by an entropy function~$S$ and a dissipation potential~$\Psi^*$.
  
  As elaborated in (I), following \cite{MPR14}, the dissipation potential $\Psi^*$ may be found by studying the stochastic process~$X_t$. In particular, we need to compute the following cumulant generating function (cf.~Eq.~28 in (I) and \cite[Chapter~1]{FK06}):
  \begin{equation}\label{nonlinearGenerator}
   \dfrac{2 k_B}{\tau} \ln\left\langle e^{\alpha \left(X_\tau - x\right)} \right\rangle_x .
  \end{equation}
  In this expression, the expectation is taken over all possible realizations of the stochastic process $X_t$ starting from $x$ and with time duration~$\tau$, and $\tau$ is far from the fast time scale $t_2$ and the slow time scale $t_1$:
  \begin{equation}
   t_2 \ll \tau \ll t_1 \, .
  \end{equation}
  
  The generalized FDT implies that 
  \begin{multline}\label{FDT}
   \dfrac{2 k_B}{\tau} \ln\left\langle e^{\alpha \left(X_\tau - x\right)} \right\rangle_x^\tau \approx \\
   \approx \Psi^*\!\left(x, \frac{\partial S(x)}{\partial x} + 2 k_B \alpha \right) - \Psi^*\!\left(x, \frac{\partial S(x)}{\partial x}\right) .
  \end{multline}
  This correspondence between the left-hand side (a property of the stochastic process) and the right-hand side (the structure of the most probable evolution) has its origin in the connection between large deviations for Markov processes and the generalized gradient flows that was proven in \cite{MPR14}, where it is explained how the connection, in the purely dissipative case, is based on a detailed-balance property of the stochastic process with respect to the distribution $e^{S(x)/k_B}$.
  
  In this paper we suppose that the distribution~$e^{S(x)/k_B}$ has already been sampled to find the function~$S(x)$, so that the static properties of the system are fully known. The main focus, instead, rests upon the computation of the dissipation potential~$\Psi^*$ by the formula~\eqref{FDT}, which completely characterizes the dynamics.
  
  From the practical standpoint, it is convenient to evaluate the expression \eqref{FDT}, for fixed $x$, at the values
  \begin{equation}
   \alpha = \dfrac{1}{2 k_B} \left( \xi - \frac{\partial S(x)}{\partial x} \right) .
  \end{equation}
  Indeed, for $\xi = 0$, we get the second term on the left-hand side of Eq.~\eqref{FDT} because $\Psi^*(x, 0) = 0$; with the other values of $\xi$ we explore the dissipation potential in $\xi$-space.
  
  It is not the purpose of this paper to construct efficient simulations, nor to pursue any statistical rigor, which we reserve for future work. For instance, in the same spirit of the inference of the infinitesimal generator of a continuous-time Markov process, a theory of inference for the nonlinear generator, the limit of the cumulant generating function~\eqref{nonlinearGenerator} as $\tau \to 0$, should be developed.
  
  \smallskip
  
  We have seen that the computation of the cumulant generating function for the stochastic process~$X_t$, the left-hand side of Eq.~\eqref{FDT}, together with the information on the static distribution, provides us with the dissipative structure, expressed in terms of an entropy function~$S$ and a dissipation potential~$\Psi^*$, for the macroscopic phenomenological equation. Note, in particular, that we do not need to assume the nature of the process~$X_t$, except that it is Markovian and it satisfies detailed balance: the Gaussian case gives rise to a quadratic dissipation potential, thus to a friction matrix. The procedure based on the generalized FDT, thus, gives a unified way of dealing with both Gaussian and non-Gaussian Markovian fluctuations, and with both the Green-Kubo and the Kramers pictures.
  
\section{Example: a simple chemical reaction}\label{sec:Kramers}
 To test the method just proposed, we have chosen the simplest example where a jump process arises from coarse-graining: the Kramers escape problem over an energy barrier \cite{haK40}, which is a model for the unimolecular chemical reaction $A \leftrightarrows B$.
 
 The dynamics of chemical reactions are described by phenomenological \emph{rate laws} \cite{pM94}. A famous example is the \emph{reaction rate equation} (RRE) \footnote{also known as \emph{law of mass action} \cite{sS87,mG12} or \emph{Guldberg-Waage dynamics} \cite{GW867}}, which we consider here in the very simple version of the unimolecular reaction $A \leftrightarrows B$. Our aim is to determine the dissipation potential of the rate law from the corresponding microscopic model, the overdamped Langevin dynamics of $n$ independent particles in a double-well potential.
 
 We present the argument in three subsections. In Sec.~\ref{A} we introduce the multiscale setup and compute the dissipation potential analytically. Since the macroscopic, reactive, system is not of the diffusive type, as we show in Sec.~\ref{Green-Kubo}, it is clear that, in the framework of this series of papers, only the generalized FDT can provide the correct dissipation potential. In Sec.~\ref{numerics} we describe the algorithm by which we compute the dissipation potential numerically.
 
 \subsection{A multi-scale view: the levels of description}\label{A}
  
  \subsubsection{The microscopic level: diffusion in a double-well potential}
   \begin{figure}[h]
   	\centering
   	\begin{tikzpicture}[scale=1.85]
   	 \draw[->] 	(-1.5,0) -- (1.5,0)					node[right] 			{$y$};
   	 \foreach \x/\xtext in {0/0, -1/-L, 1/L}
   	 \draw[shift={(\x,0)}] (0pt,2pt) -- (0pt,-2pt) 	node[below] 			{$\xtext$};
   	 \draw[->] 	(0,0) -- (0,2) 						node[right, pos=.55]	{$\Delta V$};
   	 \draw[domain=-1.5:1.5,smooth,variable=\x,black,thick] plot ({\x},{\x*\x*\x*\x - 2*\x*\x + 1}) node[above] {$V(y)$};
   	\end{tikzpicture}
   	\caption{The double-well potential $V$ that we use in the numerical experiment. The regions $A = (-\infty, 0)$ and $B = (0, \infty)$ are representative of the two chemical states, and the height of the barrier is $\Delta V \coloneqq V(0) - V(L)$.}
   	\label{fig:potential}
   \end{figure}
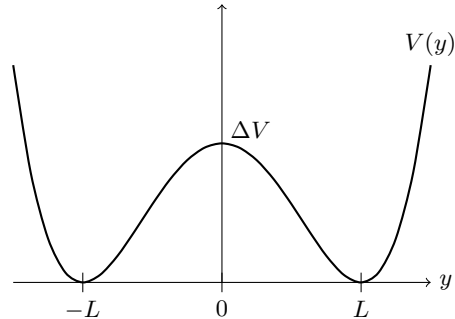
   We consider a large number~$n$ of reactive particles in a mixture. The constituents of the mixture do not directly enter our description of the system, which is based only on the state of the reactive particles. We assume that the reactive particles are independent, and the state of each particle is described by its position on the real line, which can be interpreted as a \emph{reaction coordinate} \cite[p.~1158]{pM94} or \emph{collective variable} or \emph{coordinate} \cite{LP02,ZHS16}. We gather all positions in the array $\left( y^1, y^2, \ldots, y^n \right) \in \mathbb{R}^n$.
   
   Each particle follows an overdamped Langevin dynamics in the energy landscape of \figurename~\ref{fig:potential}; in this dynamics, the noise represents the effective interaction of each particle with all constituents in the mixture, which is physically described as a heat bath at temperature $T$. The two wells of the energy landscape correspond to the two chemical states~$A$ and~$B$, and the motion is described by the SDEs
   \begin{equation}\label{Kramers-SDE}
    d Y^i_t = - \dfrac{1}{\gamma} V'(Y^i_t) \, d t + \sqrt{\dfrac{k_B T}{\gamma}} \, d W^i_t \quad\! (i = 1, \ldots, n) \, ,
   \end{equation}
   where $\gamma$ is a friction coefficient with dimensions of [mass]/[time], $V'$ is the derivative of the energy, and the $W^i$ are independent Wiener processes.
   
   Although Eq.~\eqref{diffusion} and Eq.~\eqref{Kramers-SDE} are formally equivalent, their physical roles should be clearly distinguished: Eq.~\eqref{diffusion} represents the evolution equation of a macroscopic state variable with noise enhancement in the statistical-mechanical setting of this paper; Eq.~\eqref{Kramers-SDE}, instead, is an effective microscopic dynamics.
  
  \subsubsection{The macroscopic system}
   Denoting by $\mathbbm{1}_J$ the indicator function of the set $J$, we introduce the macroscopic variable
   \begin{equation}\label{empiricalProcess}
    X(Y^1, \ldots, Y^n) := \dfrac{1}{n} \sum_{i = 1}^{n} \mathbbm{1}_B(Y^i) \, ,
   \end{equation}
   which keeps track of the concentration of the particles that, at each time $t$, are in the well $B \coloneqq (0, \infty)$; namely, it is a rational number $x$ in the set
   \begin{equation*}
    \mathcal{X}^n = \left\{0, \dfrac{1}{n}, \dfrac{2}{n}, \ldots, 1 \right\} \subset [0, 1] \eqqcolon \mathcal{X} \, .
   \end{equation*}
   The concentration of $A$ is $1 - X$, of course.
   
   From the symmetry of the problem---the states $A$ and $B$ are interchangeable---it is clear that the static distribution is a (scaled) binomial one with parameters $n$ and $1/2$,
   \begin{equation}\label{staticDistribution}
    \pi(x) = \begin{pmatrix} n \\ n x \end{pmatrix} 2^{-n} \, .
   \end{equation}
   By using Stirling's approximation, in the limit $n \to \infty$, one may verify that the distribution is in the form $e^{S(x)/k_B}$ with
   \begin{equation}\label{entropy}
    S(x) = - n k_B \left[ x \ln(2 x) + \left( 1 - x \right) \ln\!\big( 2 (1 - x) \big) \right] .
   \end{equation}
   
   We now turn to the dynamics. As one can infer from \figurename~\ref{fig:potential}, when the thermal energy~$k_B T$ is sufficiently small with respect to the height of the energy barrier~$\Delta V$, each particle spends most of its time in the minima of the potential, and the motion between the two wells happens through rapid transitions due to rare occurrences of multiple Brownian kicks in the same direction. In this regime, the system is characterized by two neatly separated time scales: the equilibration time $t_2$ of the particles in the wells, and the escape time $t_1$ from the wells \cite{DLLN16}, with $t_1 \gg t_2$. During the equilibration time $t_2$, a particle equilibrates locally and forgets where it was before the last jump event. The escape time $t_1$ defines the \emph{reaction constant} by the relation $k \coloneqq t_1^{-1}$.
   
   The stochastic process \eqref{empiricalProcess} in the low-temperature limit is the unimolecular version of the \emph{chemical master equation} (CME), a continuous-time Markov jump process that, in general, describes the evolution of the concentrations of multiple chemical species in a mixture \cite{AMPSV12}. However, since the ratio between the temperature and the energy barrier is fixed by the physics of the system and should be considered as finite, the separation of time scales is also finite, and we must think of the CME only as a Markovian approximation of the true, non-Markovian dynamics. The approximation is valid at time scales that are larger than the microscopic equilibration time scale~$t_2$: as a consequence, only correlation functions with time differences~$\tau \gg t_2$ have a true macroscopic meaning.
   
   The infinitesimal generator of the CME for our two-state system reads
   \begin{align}\label{generator}
    (\mathcal{Q} f)(x) = n k \left(1 - x\right) &\left[ f\!\left(x + \dfrac{1}{n}\right) - f(x) \right] + \nonumber \\
    + n k x &\left[ f\!\left(x - \dfrac{1}{n}\right) - f(x) \right] .
   \end{align}
   The CME is in detailed balance \cite{bJ15} with respect to the distribution \eqref{staticDistribution}. To find the dissipation potential, we compute the cumulant generating function
   \begin{align}\label{nonlinearGeneratorCME}
    &\dfrac{2 k_B}{\tau} \ln\left\langle e^{\alpha \left(X_\tau - x\right)} \right\rangle_x^\tau \stackrel{(*)}\approx \nonumber \\
    &= \dfrac{2 k_B}{\tau} \ln\left\{ 1 + \tau \left[ n k (1 - x) \left( e^{\frac{\alpha}{n}} - 1 \right) + \right. \right. \nonumber \\
    &\left. \left. \hspace{43.2mm} + \, n k x \left( e^{-\frac{\alpha}{n}} - 1 \right) \right] \right\}  \nonumber \\
    &\overset{(**)}{\approx} 2 n k_B \left[ k (1 - x) \left( e^{\frac{\alpha}{n}} - 1 \right) + k x \left( e^{- \frac{\alpha}{n}} - 1 \right) \right] .
   \end{align}   
   Then, a short computation shows that the function
   \begin{equation}\label{dissipationPotentialCME}
    \Psi^*(x, \xi) = 4 n k_B \, k \sqrt{x (1 - x)} \left( \cosh\dfrac{\xi}{2 n k_B} - 1\right)
   \end{equation}
   satisfies the defining relation \eqref{FDT}.
   \smallskip
   
   In the approximations $(*)$ and $(**)$ we have used a short-time assumption $n k \tau \ll 1$, which corresponds to the fact that, during the time scale~$\tau$, much less then one reaction event occurs, on average. By a slightly different argument one finds the same result but under the weaker condition $k\tau \ll 1$, as follows.
   
   When $k\tau \ll 1$, the stochastic process \eqref{empiricalProcess} may also be approximated by a discrete-time Markov jump process with transitions of the type \cite{dtG01}
   \begin{equation}\label{tau-leaping}
    X_t \overset{\tau}{\longmapsto} X_{t + \tau} = X_t - \dfrac{L^+_{x, \tau}}{n} + \dfrac{L^-_{x, \tau}}{n} \, ,
   \end{equation}
   where $L^+_{x, \tau}$ and $L^-_{x, \tau}$ are independent Poisson random variables with parameters
   \begin{equation*}
    \lambda^+_{x, \tau} = k \, n (1 - x) \, \tau \qquad \text{and} \qquad \lambda^-_{x, \tau} = k \, n x \, \tau \, ,
   \end{equation*}
   and corresponding probability mass functions
   \begin{align*}
    p^+_{x, \tau}(l^+) &= \dfrac{\left( \lambda^+_{x, \tau} \right)^{l^+} e^{- \lambda^+_{x, \tau}}}{l^+!} \quad \text{and} \\
    p^-_{x, \tau}(l^-) &= \dfrac{\left( \lambda^-_{x, \tau} \right)^{l^-} e^{- \lambda^-_{x, \tau}}}{l^-!} \, .
   \end{align*}
   
   The calculation
   \begin{align}\label{discrete}
    & \dfrac{2 k_B}{\tau} \ln\left\langle e^{\alpha \left(X_\tau - x\right)} \right\rangle_x^\tau = \nonumber \\
    &= \dfrac{2 k_B}{\tau} \ln\sum\limits_{l^+, \, l^-} e^{\alpha \frac{l^- - l^+}{n}} p^+_{x, \tau}(l^+) \, p^-_{x, \tau}(l^-) = \nonumber \\
    &= 2 n k_B \left[ k \, (1 - x) \left(e^{\frac{\alpha}{n}} -1\right) + k \, x \left(e^{-\frac{\alpha}{n}} -1\right) \right]
   \end{align}
   then produces the same dissipation potential \eqref{dissipationPotentialCME}. Note that, under the stronger assumption $n k \tau \ll 1$, the occurrences $l^+, l^- > 1$ carry only negligible weight in \eqref{discrete}.
   
  \subsubsection{The most probable evolution: the reaction rate equation}
   As one may infer from the generator \eqref{generator}, when $n \to \infty$, the CME converges to the macroscopic phenomenological equation
   \begin{equation}\label{RRE}
    \dfrac{d x_t}{d t} = k \left( 1 - 2 x_t \right) ,
   \end{equation}
   where $x \in \mathcal{X} = [0, 1]$. With the entropy function \eqref{entropy} and the dissipation potential \eqref{dissipationPotentialCME}, one may verify that the \emph{reaction-rate equation} (RRE) \eqref{RRE} can be written as the generalized gradient flow
   \begin{equation}
    \dfrac{d x_t}{d t} = \left. \frac{\partial \Psi^*(x_t, \xi_t)}{\partial \xi} \right\vert_{\xi_t = \tfrac{\partial S(x_t)}{\partial x}} .
   \end{equation}
   
   The generalized gradient flow given by the pair \eqref{entropy}, \eqref{dissipationPotentialCME} was proposed in \cite{mG93} (see also \cite{mG10}) independently from any consideration about an underlying stochastic process, which is remarkable in view of the fact that the number of candidate dissipation potentials, given the same entropy \eqref{entropy}, is infinite (cf.~\cite[Example~4.3]{aM19}). For instance, a quadratic dissipation potential was proposed in \cite{OG97,aM11} and was supported in \cite{hcO15} by thermodynamic and geometric arguments. The corresponding friction matrix reads
   \begin{equation}\label{M} 
    M(x) = \dfrac{k}{n k_B} \Lambda(1 - x, x) \, ,
   \end{equation}
   with the logarithmic mean
   \begin{equation}
    \Lambda(x, y) \coloneqq
     \begin{dcases}
      x 						& x = y \, , \\
      \dfrac{x - y}{\ln x - \ln y} 	& x \neq y \, .
     \end{dcases}
   \end{equation}
   
   We elaborate further on the dissipation potential \eqref{dissipationPotentialCME} in Sec.~\ref{numerics}, where we show that the same can be obtained by numerical simulations of the microscopic system, with the only assumptions that the macroscopic stochastic process is Markovian and in detailed balance. In the next section, instead, we see what would happen if we restricted the class of stochastic processes to diffusions.
   
 \subsection{Green-Kubo formula and the shortcomings of diffusion approximations}\label{Green-Kubo}
  The generalized FDT provides us with the dissipation potential and the entropy function of the RRE from the properties of the associated stochastic process. Specifically, it gives us (i) an entropy function that can be sampled from the static distribution and (ii) a dissipation potential that can be sampled from the cumulant generating function; moreover, (iii) the generalized gradient flow constructed from the entropy and the dissipation potential produces the correct form of the RRE. Can these three features be reproduced by assuming the stochastic process~$X_t$ to be a diffusion process? In this section we find an answer.
  
  Note that the assumption of a diffusion process is not just an academic exercise since, in practical situations, we do not know the nature of the macroscopic process~$X_t$, but only have access to the dynamics of a more microscopic model. Assuming a diffusion process means restricting ourselves to the classical scheme of Green-Kubo relations.
  
  \smallskip
  
\begin{figure}
   \centering
   \includegraphics[width=.9\linewidth]{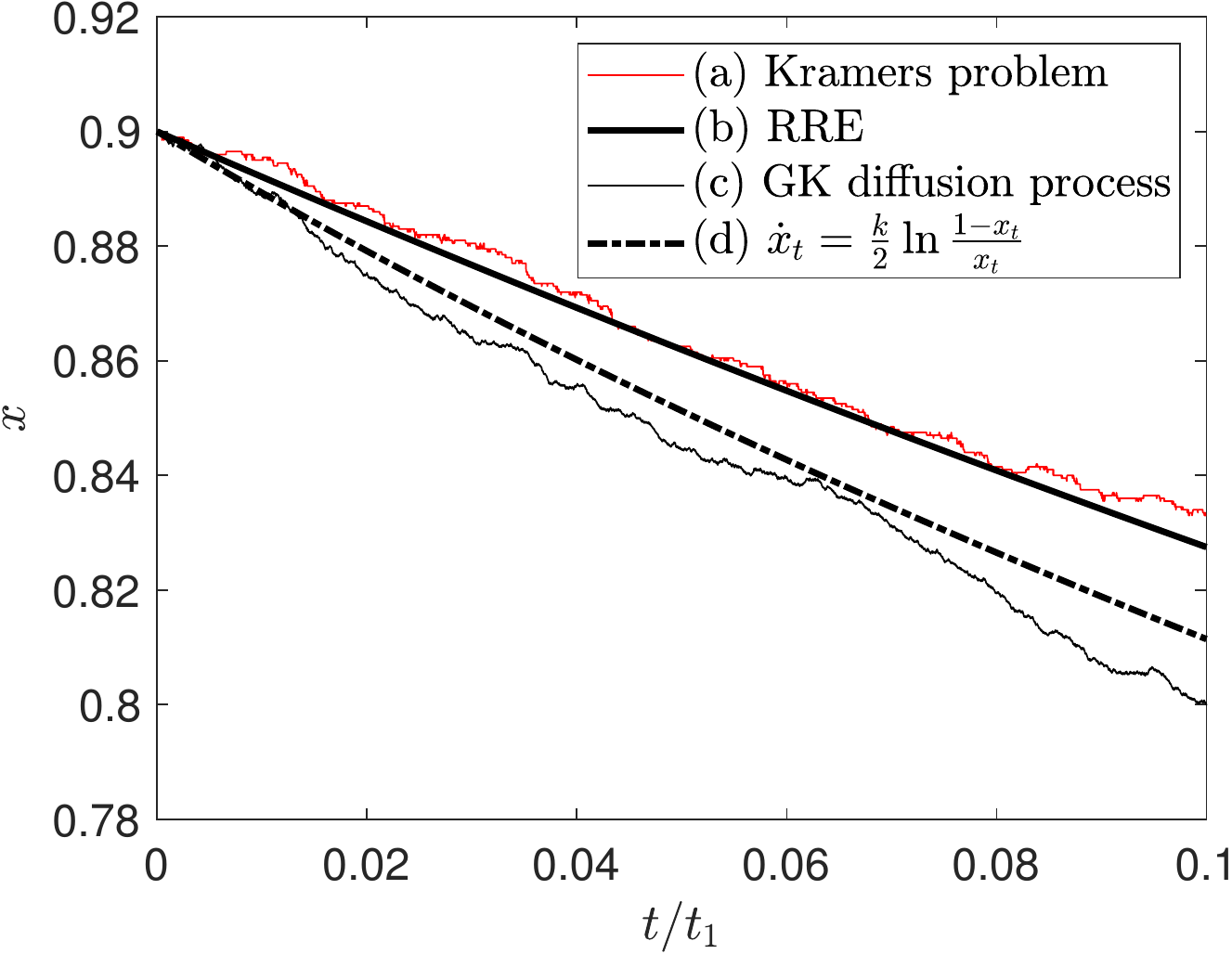}
   \caption{Comparison between the correct most probable evolution, given by the RRE, and the most probable evolution of the `Green-Kubo' diffusion process. (a) Simulation of the true Kramers problem and average over 200 realizations. (b) Analytical solution of the RRE. (c) Simulation of the `Green-Kubo' diffusion process and average over 200 realizations (10 realizations have fallen out of the domain $[0, 1]$). (d) Numerical solution of the most probable evolution corresponding to the diffusion process `Green-Kubo'. The values of the parameters are: $\Delta V/(k_B T) = 4$, $\Delta t = 10^{-2}$, $n = 10$, and $k = \bar{k}$ according to the Kramers formula \eqref{Kramers}.}
   \label{fig:GK}
  \end{figure}

  Let us assume that the `Green-Kubo' diffusion process
  \begin{equation*}
   d X_t = M_\text{GK}(X_t) S'(X_t) \, d t + \sqrt{2 k_B M_\text{GK}(X_t)} \diamond d W_t
  \end{equation*}
  is an accurate dynamics for the macroscopic variables. Since it is a diffusion, the corresponding dissipation potential is quadratic, and the friction matrix may be computed by the Green-Kubo formula
   \begin{equation}
    2 k_B M_\text{GK}(x) = \lim\limits_{\tau \to 0} \dfrac{1}{\tau} \mathbb{E}\!\left[ \left( X_\tau - x \right)^2 \Big| X_0 = x \right] .
   \end{equation}
   Note
  that, in practice, the actual sampling is made with simulations at a more microscopic level and, since in this case we know that the CME is the correct dynamics for the macroscopic variables, from Eq.~\eqref{GKgenerator} we expect the result
  \begin{align}
   2 k_B M_\text{GK}(x) &= \left[ (\mathcal{Q} f)(x) - 2 x (\mathcal{Q} g)(x) \right] = \nonumber \\
   &= \dfrac{k \left( 1 - x \right) + k x}{n} = \dfrac{k}{n} \label{M-CLE} \, ,
  \end{align}
  where $f(x) = x^2$, $g(x) = x$, and $\mathcal{Q}$ is the generator \eqref{generator} of the CME. We remark that $M_\text{GK}$ is, in general, a function of $x$, and it would be so if there were two distinct rate constants $k^- \neq k^+$ for the backward and forward reactions.
  With the entropy \eqref{entropy}, we get
  \begin{equation*}
   d X_t = \dfrac{k}{2} \ln\dfrac{1 - X_t}{X_t} \, d t + \sqrt{\dfrac{k}{n}} \, d W_t \, ,
  \end{equation*}
  which has (i) the correct stationary distribution, that is, the correct entropy (by construction), (ii) a Green-Kubo expression for the friction matrix that can be computed by simulations, but (iii) the wrong drift and, therefore, the wrong most probable evolution (cf.~\figurename~\ref{fig:GK}). (Note that while `spurious' drift terms sometimes appear as the result of the different possible interpretations of the noise in an SDE, in this case the noise is additive, and no such terms arise.) 
  
  \smallskip
  
\begin{figure}
   \centering
   \includegraphics[width=.835\linewidth]{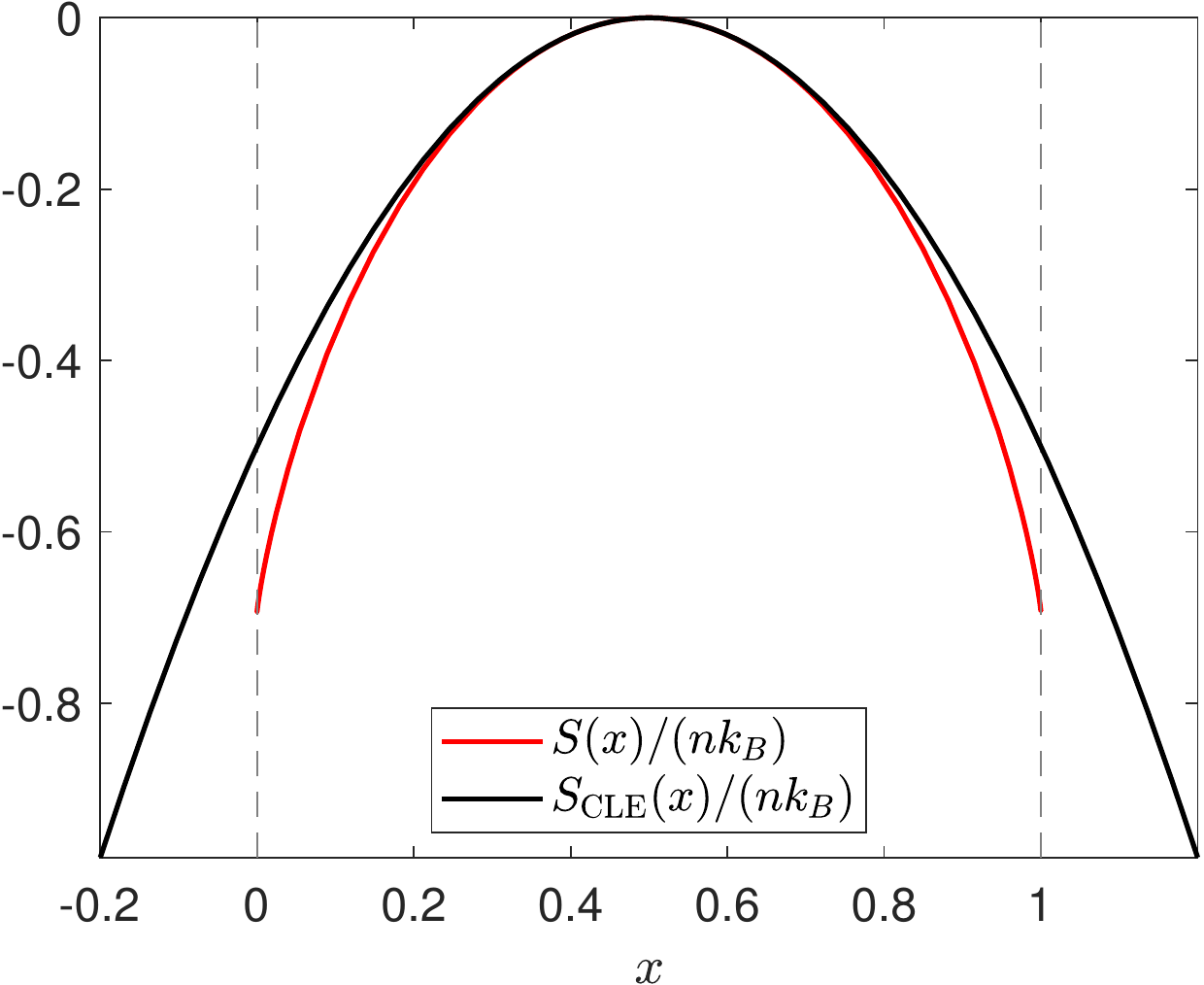}
   \caption{Comparison between the entropy functions associated to the stationary distribution of the CME and the CLE. The stationary distribution~$e^{S_\text{CLE}/k_B}$ implies a nonzero probability for states outside the domain~$[0, 1]$.}
   \label{fig:entropy}
  \end{figure}

  Other possibilities of constructing a diffusion process for chemical reactions have been proposed: their goal is to approximate the CME for a large number of particles. One of them is called the \emph{chemical Langevin equation} \cite{dtG00}, which can also be interpreted as the diffusion approximation \footnote{The diffusion approximation may be found by three heuristic arguments: the first corresponds to expanding the generator to first order in~$1/n$, the second to considering the Kramers-Moyal expansion in the equation for the law, and the third to replacing the Poisson noise, which describes the jumps, by a Brownian one for large~$n$ \cite{ngvK83,EK05,AK11}.} of the CME, and reads
  \begin{align}
   &d X_t = k \left(1 - 2 X_t\right) \, d t + \sqrt{\dfrac{k}{n}} \, d W_t = \label{CLE} \\
   &= M_\text{GK}(X_t) S'_\text{CLE}(X_t) \, d t + \sqrt{2 k_B M_\text{GK}(X_t)} \diamond d W_t \, , \nonumber
  \end{align}
  with the entropy
  \begin{equation}
   S_\text{CLE}(x) = - 2 n k_B \left( x - \dfrac{1}{2} \right)^2
  \end{equation}
  and the friction matrix \eqref{M-CLE}. This process has (iii) the correct most probable evolution, (ii) a Green-Kubo expression for the friction matrix, but (i) the wrong entropy, viz., the wrong stationary distribution (cf.~\figurename~\ref{fig:entropy}).
  
  \smallskip
  
  A third choice is the \emph{log-mean equation} \cite{BBGBD15}
  \begin{multline*}
   d X_t = M(X_t) S'(X_t) \, d t + \sqrt{2 k_B M(X_t)} \diamond d W_t = \\
   = k \left(1 - 2 X_t\right) \, d t + \sqrt{\dfrac{2 k}{n} \Lambda(1 - X_t, X_t)} \diamond d W_t \, ,
  \end{multline*}
  with the entropy \eqref{entropy} and the friction matrix \eqref{M}. It has (iii) the correct most probable evolution, (i) the correct stationary distribution, but (ii) the friction matrix cannot be sampled by the Green-Kubo formula (cf.~later \figurename~\ref{fig:fitting}).
  
  A common feature of all three diffusion processes is that the macroscopic variable $X$ has a positive probability of exiting the domain $[0, 1]$, and this phenomenon may be seen both in the stationary distribution (cf.~\figurename~\ref{fig:entropy}) and in the dynamics (in the simulation behind \figurename~\ref{fig:GK}, 10 out of 200 realizations of the `Green-Kubo' diffusion process have exited the domain $[0, 1]$).
  
  \smallskip
  
  {\setlength{\tabcolsep}{5.85pt}
   \def\arraystretch{1.6}%
   \begin{table}
    \centering
  	\begin{tabular}{rccc}%
  	 & GK & CLE & LME \\[\jot]
  	 $S$ can be sampled & \cmark & \xmark & \cmark \\
  	 $M$ can be sampled & \cmark & \cmark & \xmark \\
  	 $\dot x = M(x) \frac{\partial S(x)}{\partial x}$ is correct evolution & \xmark & \cmark &\cmark
  	\end{tabular}
    \caption{None of the three Gaussian-based methods studied in Sec.~\ref{Green-Kubo} performs correctly on all three criteria.}%
  	\label{table:comparison}%
   \end{table}}
  In conclusion, there is no way for a diffusion process to satisfy the requirements (i)-(iii) simultaneously (cf.~\tablename~\ref{table:comparison}). In order to satisfy all three requirements, one should move to more general Markov processes and to non-quadratic dissipation potentials.

 \subsection{Numerical experiments}\label{numerics}
  The goal of the coarse-graining scheme proposed here is to infer the structure of a more macroscopic level of description from a more microscopic one, which in this case is represented by the overdamped Langevin dynamics of many independent particles in a double-well potential. In such a simple situation, the macroscopic variable is easily chosen as the concentration of particles in one well. The dynamics of the concentration, when the thermal energy~$k_B T$ is small with respect to the energy barrier~$\Delta V$, is well approximated by a Markov process. However, we stress the fact that, in general, the system over which we have full control is the microscopic one; even if, in this simple case, we know the form of the macroscopic Markov process and the dissipation potential (up to the parameter~$k$ that completely characterizes them), in the numerical calculations we pretend to have no macroscopic information.
  
  Before describing our numerical method, and to highlight the free parameters of the system, let us make Eq.~\eqref{Kramers-SDE} dimensionless through the friction coefficient~$\gamma$, the energy barrier~$\Delta V$, and $L$, the distance between the maximum and the minimum points in the energy landscape. Defining by $\tilde{a}$ the dimensionless quantity relative to $a$, we find
  \begin{equation}
  d \widetilde{Y}^i_{\tilde{t}} = - \widetilde{V}'\big(\widetilde{Y}^i_{\tilde{t}}\big) \, d \tilde{t} + \sqrt{\dfrac{k_B T}{\Delta V}} \, d \widetilde{W}^i_{\tilde{t}} \quad (i = 1, \ldots, n) \, .
  \end{equation}
  From the numerical calculations and the values of the dimensional parameters $\gamma$, $\Delta V$, and $L$, one may find the physically meaningful results. To simplify the notation, from here on we drop the tildes.
  
  The method advanced in this paper aims at computing the dissipation potential numerically. When, like in this case, we know its functional form, we could also determine the value of its unknown parameter~$k$ and compare it with the reference one given by Kramers' formula \cite{haK40,BdH15}
  \begin{equation}\label{Kramers}
   \bar{k} = \dfrac{\sqrt{- V''(0) \, V''(L)}}{2 \pi} e^{- 2 \, \Delta V/(k_B T)} .
  \end{equation}
  
  In accordance with the approach of this paper, we suppose that the entropy function~$S$, or its derivative $\partial S/\partial x$ \cite{IOK09}, has already been found, and we concentrate on the dynamics by computing the cumulant generating function \eqref{nonlinearGenerator}. The evaluation of the cumulant generating function depends on the energy landscape, which we choose as
  \begin{equation*}
   V(y) = \left( y^2 - 1 \right)^2 ,
  \end{equation*}
  and on five parameters:
  \begin{itemize}
   \item The ratio~$r \coloneqq \Delta V/(k_B T)$, which controls the separation of time scales. Kramers mentions in his seminal paper \cite{haK40} that $r = 2.5$ is sufficient: the process becomes approximately Markov. We actually work with $r = 4$, which implies $\bar{k} \approx 4 \cdot 10^{-5}$. In real applications, however, this is not a parameter, but a model datum.
   \item The time-step size $\Delta t \to 0$ of the numerical scheme used to simulate the SDE \eqref{Kramers-SDE}, which should resolve the microscopic dynamics, guarantee stability of the scheme, and be smaller than the local equilibration time~$t_2$~\footnote{An estimation of the equilibration time~$t_2$ should be done separately, for instance with the help of the various methods available in the literature \cite{DLLN16,afV98,BLS15}. In our case, the ratio $t_2/\Delta t$ is of the order of $20$ \cite[Figure~6.4]{aM19}.}. We consider the numerical value $\Delta t = 0.01$.
   \item The time interval $\tau$. This time constant should be ``macroscopically small'', i.e., much smaller than the typical jump time $t_1 = k^{-1}$, but also larger than the equilibration time $t_2$, in such a way that we retain only the macroscopic features of the process~$X_t$. We take $\tau = \bar{k}^{-1}/50 \approx 5 \cdot 10^2$, where $\bar{k}$ is given by formula \eqref{Kramers}. In more general contexts, we may not know the values of the characteristic times in advance and should perform an appropriate estimation of them.
   \item The number of particles $n \to \infty$. Since the particles are independent, the cumulant generating function factorizes into cumulant generating functions for the single particles. Its computation would then require the simulation of just one particle. However, since $n$ controls the discretization of the space $\mathcal{X}^n$, and to keep the generality of our framework, which should work for interacting particle systems as well, we choose $n = 10$.
   
   \item The sample size $N \to \infty$ over which we calculate the empirical expectation. To obtain good statistics, sufficiently many jumps should occur in the total observation time. Since the average jump time is $t_1$, we need $N \tau \gg t_1$. With the choice $N = 10^5$, the total average number of jumps is $N \tau/t_1 \approx 2 \cdot 10^3$.
  \end{itemize}
  We have thus built the chain of inequalities
  \begin{equation*}
   \Delta t \ll t_2 \ll \tau \ll t_1 \ll N \tau \, .
  \end{equation*}
  
  \smallskip
  
  The numerical setup is the following.
  \begin{enumerate}
   \item For every $x \in \mathcal{X}^n$, we select a series of values $\alpha_j = (\xi_j - \partial S/\partial x)/(2 k_B)$, with the $\xi_j$ logarithmically spaced \footnote{We choose a small value $\xi^+_1$. The positive values $\xi^+_j$ are in geometric progression starting from $\xi^+_1$, and the negative ones are $\xi^-_j = - \xi^+_j$.} in the interval $[-\xi_\text{max}, \xi_\text{max}]$ with $\xi_\text{max} = 8$. The logarithmic spacing has the aim of resolving the region around $\xi = 0$ in a sufficiently accurate way.
   \item For every $x$, we run $N$ simulations, of length $\tau$, of the $n$ independent SDEs \eqref{Kramers-SDE}. We use an Euler-Mayurama scheme with step size $\Delta t$. The starting point for $n x$ particles is $1$ and for the others is $-1$, that is, the initial microscopic state is
   \begin{equation*}
    ( \underbrace{1, 1, \ldots, 1}_{n x}, \underbrace{-1, -1, \ldots, -1}_{n - n x} ) \, .
   \end{equation*}We need not care about equilibration in the wells because $\tau \gg t_2$.
   \item We index the simulations by $\ell$ and say that the random variable $X_\tau$ has $x^\ell_\tau$ as its realization. After the $\ell$-th simulation started from $x$, we compute the quantities
   \begin{equation}\label{sample}
    h_j^\ell(x) := e^{\alpha_j \left(x^\ell_\tau - x\right)} \, .
   \end{equation}
   There is one such quantity for each $x$, $j$ and $\ell$.
   \item To estimate the expectation in Eq.~\eqref{nonlinearGenerator}, we calculate
   \begin{equation}
    \dfrac{2 k_B}{\tau} \ln\bigg( \dfrac{1}{N} \sum\limits_{\ell = 1}^N h_j^\ell(x) \bigg) \, .
   \end{equation}
  \end{enumerate}
  From the cumulant generating function we obtain an estimate of the dissipation potential~$\widehat{\Psi}^*(x, \xi)$ by using the central equation \eqref{FDT}.
  
  \begin{figure}
    \centering
    \includegraphics[width=\linewidth]{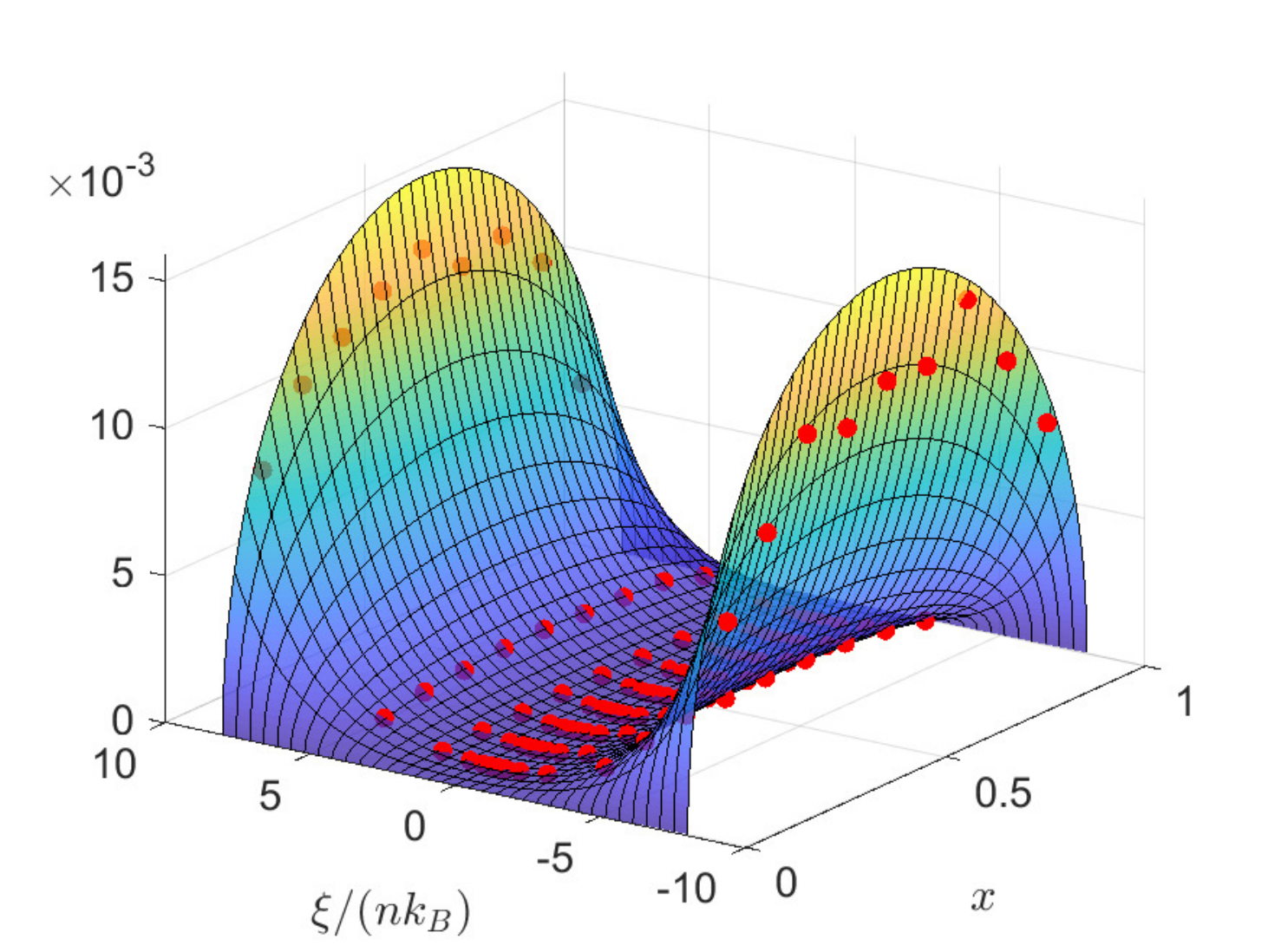}
  	\caption{Numerical estimate of the dissipation potential $\widehat{\Psi}^*/(n k_B)$ compared with the reference result. The latter is represented by the smooth surface, and the former by the red dots. The values of the parameters are: $r = 4$, $\Delta t = 10^{-2}$, $\tau = \bar{k}^{-1}/200$, $n = 10$, $N = 10^5$, $\xi_\text{max} = 8$.}
  	\label{fig:dissipationPotential}
  \end{figure}
  \begin{figure}
  	\includegraphics[width=.818\linewidth]{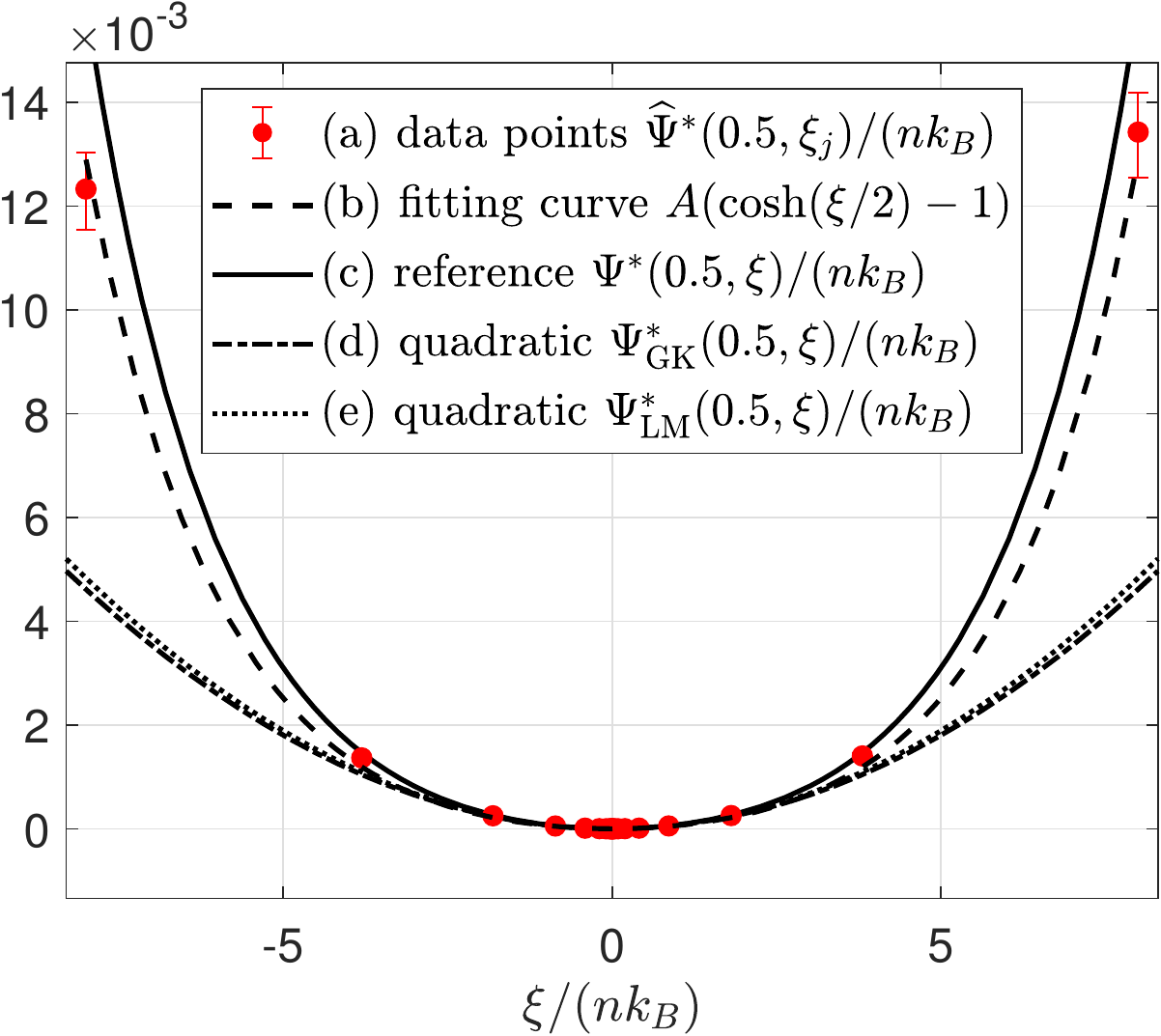}\vspace{-7.5mm}
  	\includegraphics[width=.818\linewidth]{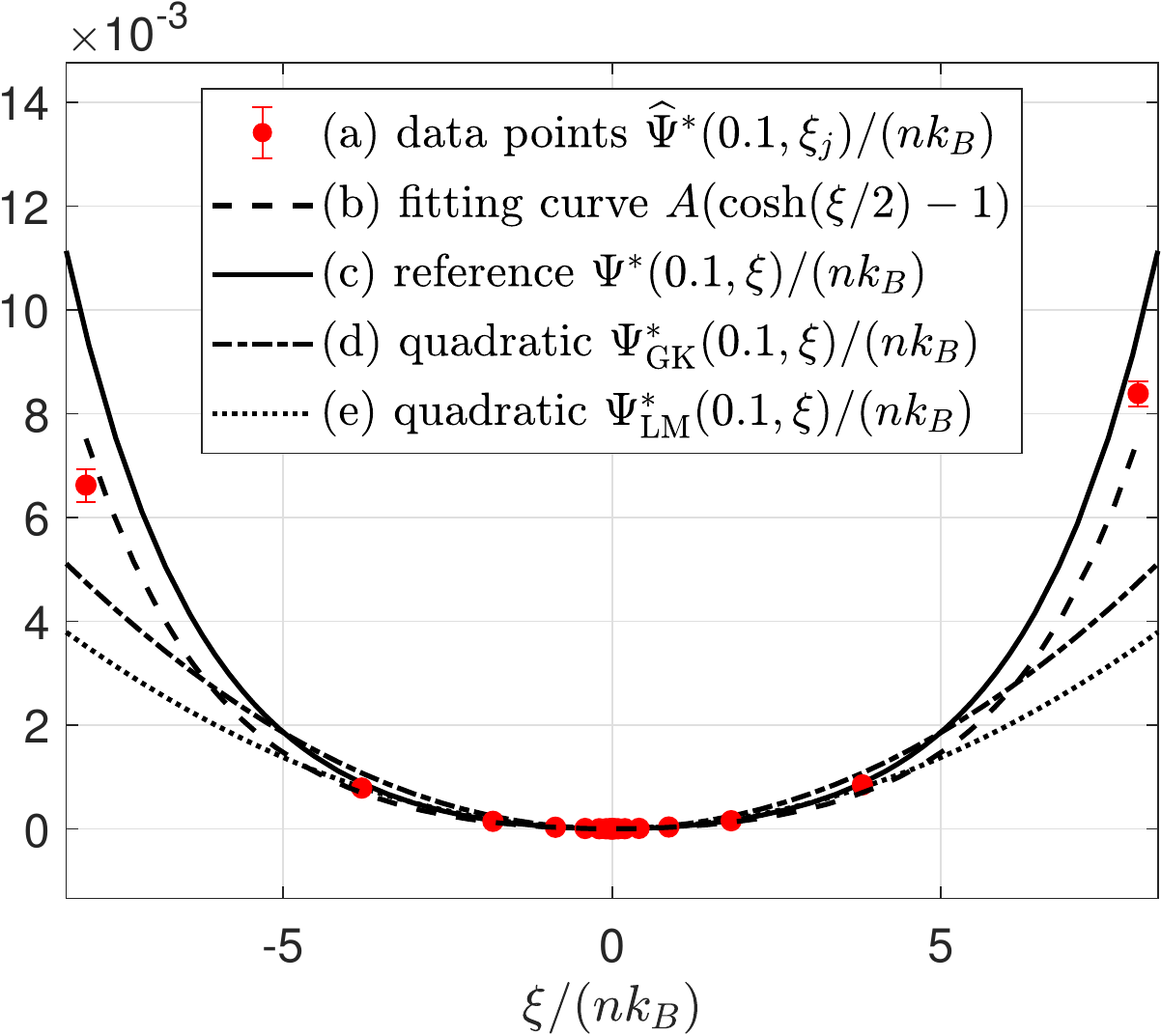}
  	\caption{(a) The data points $\widehat{\Psi}^*(x, \xi)/(n k_B)$ ($x = 0.5, 0.1$) are fitted with (b) the function $A \left( \cosh(\xi/2) - 1 \right)$ (cf.~Eq.~\eqref{dissipationPotentialCME}) and compared with (c) the reference dissipation potential with $k = \bar{k}$. The confidence intervals are at level 99\%. In addition, we compare (d) the dissipation potential~\eqref{PsiGK} derived from the friction matrix \eqref{M-CLE}, which can be computed by the Green-Kubo formula, with (e) the dissipation potential~\eqref{PsiLM} derived from the friction matrix~\eqref{M} with $k = \bar{k}$. The two agree only near the equilibrium point $x = 0.5$.}
  	\label{fig:fitting}
  \end{figure}
  
  The results of the algorithm are displayed in \figurename~\ref{fig:dissipationPotential}, where the solid surface is the reference dissipation potential \eqref{dissipationPotentialCME} with ${k = \bar{k}}$, and the red dots are its estimated values, which show good agreement. If we assume the functional form of the dissipation potential given by Eq.~\eqref{dissipationPotentialCME}, we may determine the reaction constant by fitting the simulation points with the function \eqref{dissipationPotentialCME}. We do so with the values of~$\widehat{\Psi}^*(x, \xi)$ at the points~$x = 0.5, 0.1$. In \figurename~\ref{fig:fitting}, we compare this fitting procedure to the reference dissipation potential with $k = \bar{k}$. We observe a small discrepancy, which we do not fully understand, but we do not know a simple alternative to compute the dissipation potential numerically. We only mention that the Kramers formula overestimates the reaction constant~$k$ for any finite $r$; see \cite{DLLN16} for a more sophisticated method to compute~$k$.
  
  \smallskip
  
  In addition, we show how the diffusion setting does not work in this example. In the previous subsection we have seen how two diffusion approaches, `Green-Kubo' and `Chemical Langevin equation', may be constructed from a friction matrix that can be computed by the Green-Kubo formula, but do not reproduce either the correct most probable evolution or the static distribution of the process~$X_t$. We now demonstrate numerically how also the third diffusion process, the `log-mean equation', is inconsistent, namely the corresponding friction matrix~\eqref{M} cannot be sampled by the Green-Kubo formula~\eqref{M-CLE}. The friction matrix~\eqref{M} would give rise to the dissipation potential
  \begin{equation}\label{PsiLM}
   \Psi_\text{LM}^*(x, \xi) = \dfrac{k}{2 n k_B} \Lambda(1 - x, x) \, \xi^2 \, ,
  \end{equation}
  whereas the Green-Kubo formula would suggest
  \begin{equation}\label{PsiGK}
   \Psi_\text{GK}^*(x, \xi) = \dfrac{k}{4 n k_B} \xi^2 \, .
  \end{equation}
  The two agree in the region around the equilibrium point $x = 0.5$, but differ significantly for more extreme values of $x$, as one observes in the second frame of~\figurename~\ref{fig:fitting}. From the expressions \eqref{dissipationPotentialCME}, \eqref{PsiLM}, and \eqref{PsiGK}, we also note that, in the same region around $x = 0.5$, all dissipation potentials agree for values of~$\xi$ close to 0: in particular, at $x = 0.5$ and close to $\xi = 0$, the dissipation potentials \eqref{PsiLM} and \eqref{PsiGK} are the second-order Taylor approximations of the dissipation potential~\eqref{dissipationPotentialCME}.
  
  \smallskip
  
  We emphasize that our aim has been to directly resolve the dissipative structure of the macroscopic equations: this is not the aim of standard approaches \cite{eVE06,FVEW15,HBSBS14}. We also expect that our viewpoint will constitute an advantageous tool for less simple systems, such as problems of homogenization~\cite{MS13} or systems where the structure of the macroscopic dynamics is not known in advance \cite{BC12}.
  
  If we consider more general reaction networks, characteristic problems of standard approaches will remain: for instance, high local minima of the potential landscape are rarely explored and boundaries between macroscopic states are not easy to set \cite{FVEW15}. The method proposed here presents the additional complication of being ``stiff'' because of the strong nonlinearity in Eq.~\eqref{sample}.
  
  To tackle the issues of numerical efficiency, \emph{importance sampling} has been established as the basis of many known numerical methods in statistical mechanics~\cite{hT11}: the probability distribution of a random variable is changed, often by ``exponential tilting'', in such a way that rare events become less rare and can more easily be observed. Physically, this corresponds to biasing the system by an external force. Following the ideas in \cite{HBSBS14}, we would like to improve our numerical algorithm by importance-sampling techniques.
  
\section{Conclusions}\label{sec:conclusions}

 The main subject of this series of papers is a generalization of the fluctuation-dissipation theorem of the second kind (FDT). In paper (I), inspired by \cite{MPR14}, we postulated this generalization on the basis of physical principles. In this second paper we have used the generalized FDT to develop a new method of coarse-graining. By this method, the dissipative structure (force-flux constitutive law) of a given phenomenological equation is expressed in terms of a dissipation potential and an entropy function, and can be computed by numerical simulations on a more microscopic level of description. We have illustrated the success of the method with the example of the simple chemical reaction $A \leftrightarrows B$ by computing the dissipation potential explicitly (cf.~\figurename~\ref{fig:dissipationPotential} and \ref{fig:fitting}). It was not our aim to address the choice of good macroscopic variables, which usually constitutes a problem-dependent challenge. Here we assume them to be given. In contrast to (I), here the theory has been restricted to purely dissipative systems.

 \bigskip
 
 We now recapitulate our findings in more detail. First, to realize why we need the generalization at all, we have studied the classical FDT for diffusion processes, its power and its limitations. The classical FDT suggests two facts: first, that the most probable paths of a diffusion process, the macroscopic phenomenological equations, are solutions of the gradient flow
 \begin{equation}
\label{eq:GF-summary}
  \dfrac{d x_t}{d t} = M(x_t) \frac{\partial S(x_t)}{\partial x} \, ,
 \end{equation}
 and secondly, that the \emph{friction} matrix $M$ above coincides with the \emph{diffusion} matrix of the process. Once we have computed the function~$S$ by studying a static problem, we can compute the friction matrix~$M$ by estimating the infinitesimal covariance matrix of the process. This is the essence of the Green-Kubo formula.
 
 We have then remarked that some thermodynamic variables have fluctuations that are not well described by diffusion processes, and the typical example is a chemical reaction, which is better represented as a jump process. In Sec.~s\ref{Green-Kubo} and \ref{numerics} (\figurename~\ref{fig:GK}, \ref{fig:entropy}, and \ref{fig:fitting}), with the example of the chemical reaction $A \leftrightarrows B$, we have shown how applying the Green-Kubo picture to these systems leads to incorrect results: there is no way of determining a consistent dissipative structure by restricting the fluctuations to the class of diffusion processes. Indeed, these systems have always been studied under the `Kramers picture', where jump transition rates are the central quantities to be computed. But this picture does not identify any force-flux constitutive law.
 
 The generalized FDT allows us to solve this problem by using the setting of general Markov processes. Like the classical FDT, the generalized FDT suggests two facts: first, the most probable paths of a Markov process with detailed balance are solutions of the generalized gradient flow
 \begin{equation}
 \label{eq:GGF-summary}	
   \dfrac{d x_t}{d t} = \left. \frac{\partial \Psi^*(x_t, \xi_t)}{\partial \xi} \right\vert_{\xi_t = \tfrac{\partial S(x_t)}{\partial x}} ,
 \end{equation}
 and secondly, the fluctuations of the process are characterized by the dissipation potential $\Psi^*$---the same dissipation potential that appears in~\eqref{eq:GGF-summary}.
  
 In this way the generalized FDT recognizes the dissipation potential $\Psi^*$ as the fundamental object that characterizes fluctuations. The classical, diffusive, case corresponds to a dissipation potential $\Psi^*$ that is quadratic, and therefore characterized as $\Psi^*(x,\xi) = \tfrac12 \xi^T M(x)\xi$; one then readily recognizes the matrix $M$ in~\eqref{eq:GF-summary} as the second derivative $\partial^2 \Psi^*/\partial\xi^2$, and the quadratic functional~$\Psi^*$ as the small-noise-limit dissipation potential for a diffusion process with diffusion matrix~$M$. This also explains why the Green-Kubo relations allow the characterization of the full potential~$\Psi^*$ by only estimating the second derivative $\partial^2 \Psi^* / \partial \xi^2$ at zero.

 In general, however, the dissipation potential $\Psi^*$ is not quadratic, and a full characterization of the fluctuations requires determining the full potential $\Psi^*$. 
 We have shown that this can be done by estimating a cumulant generating function of the process, and we have proposed a numerical method to do this. By this method we do not need to assume anything about the macroscopic stochastic process beyond  Markovianity and detailed balance, and thus we can deal with diffusion and jump processes in a common framework, which unifies quadratic and non-quadratic dissipation potentials and the Green-Kubo and the Kramers pictures.
 
 The example of a simple chemical reaction $A \leftrightarrows B$ has shown both the power of the method and the failure of the Green-Kubo schemes when applied to a jump process: the computed dissipation potential agrees well with the theoretical one (\figurename~\ref{fig:dissipationPotential} and \ref{fig:fitting}), and by neither the Green-Kubo nor any diffusion-based method it is possible to obtain a friction matrix that, combined with the correct entropy, reproduces the correct macroscopic phenomenological equation (\figurename~\ref{fig:GK}, \ref{fig:entropy}, and \ref{fig:fitting}). The small discrepancy in the dissipation potential, visible in \figurename~\ref{fig:fitting}, requires to be investigated further and highlights the need of a deeper understanding and, consequently, of better algorithms.
 
 Although this elementary illustration has been proven successful, it certainly requires refinement from both the standpoints of the statistical solidity and of the efficiency of the algorithm, especially because we intend to apply the method to more complex systems, such as plasticity, the dynamics of glasses, nucleation theory, or the Boltzmann equation. 

\begin{acknowledgments}
 We are grateful to Mohsen Talebi, Aleksandar Donev, Patrick Ilg, Robert Riggleman, and Elijah Thimsen, who considerably helped us to improve our understanding and the presentation of the ideas.
\end{acknowledgments}

\bibliographystyle{apsrev4-2}

\begin{thebibliography}{69}%
	\makeatletter
	\providecommand \@ifxundefined [1]{%
		\@ifx{#1\undefined}
	}%
	\providecommand \@ifnum [1]{%
		\ifnum #1\expandafter \@firstoftwo
		\else \expandafter \@secondoftwo
		\fi
	}%
	\providecommand \@ifx [1]{%
		\ifx #1\expandafter \@firstoftwo
		\else \expandafter \@secondoftwo
		\fi
	}%
	\providecommand \natexlab [1]{#1}%
	\providecommand \enquote  [1]{``#1''}%
	\providecommand \bibnamefont  [1]{#1}%
	\providecommand \bibfnamefont [1]{#1}%
	\providecommand \citenamefont [1]{#1}%
	\providecommand \href@noop [0]{\@secondoftwo}%
	\providecommand \href [0]{\begingroup \@sanitize@url \@href}%
	\providecommand \@href[1]{\@@startlink{#1}\@@href}%
	\providecommand \@@href[1]{\endgroup#1\@@endlink}%
	\providecommand \@sanitize@url [0]{\catcode `\\12\catcode `\$12\catcode
		`\&12\catcode `\#12\catcode `\^12\catcode `\_12\catcode `\%12\relax}%
	\providecommand \@@startlink[1]{}%
	\providecommand \@@endlink[0]{}%
	\providecommand \url  [0]{\begingroup\@sanitize@url \@url }%
	\providecommand \@url [1]{\endgroup\@href {#1}{\urlprefix }}%
	\providecommand \urlprefix  [0]{URL }%
	\providecommand \Eprint [0]{\href }%
	\providecommand \doibase [0]{https://doi.org/}%
	\providecommand \selectlanguage [0]{\@gobble}%
	\providecommand \bibinfo  [0]{\@secondoftwo}%
	\providecommand \bibfield  [0]{\@secondoftwo}%
	\providecommand \translation [1]{[#1]}%
	\providecommand \BibitemOpen [0]{}%
	\providecommand \bibitemStop [0]{}%
	\providecommand \bibitemNoStop [0]{.\EOS\space}%
	\providecommand \EOS [0]{\spacefactor3000\relax}%
	\providecommand \BibitemShut  [1]{\csname bibitem#1\endcsname}%
	\let\auto@bib@innerbib\@empty
	\bibitem [{\citenamefont {Kubo}\ \emph {et~al.}(1991)\citenamefont {Kubo},
		\citenamefont {Toda},\ and\ \citenamefont {Hashitsume}}]{KTH85}%
	\BibitemOpen
	\bibfield  {author} {\bibinfo {author} {\bibfnamefont {R.}~\bibnamefont
			{Kubo}}, \bibinfo {author} {\bibfnamefont {M.}~\bibnamefont {Toda}},\ and\
		\bibinfo {author} {\bibfnamefont {N.}~\bibnamefont {Hashitsume}},\ }\href
	{https://doi.org/10.1007/978-3-642-58244-8} {\emph {\bibinfo {title}
			{Statistical Physics II: Nonequilibrium Statistical Mechanics}}},\ \bibinfo
	{edition} {2nd}\ ed.\ (\bibinfo  {publisher} {Springer},\ \bibinfo {year}
	{1991})\BibitemShut {NoStop}%
	\bibitem [{\citenamefont {Landau}\ and\ \citenamefont {Lifshitz}(1980)}]{LL13}%
	\BibitemOpen
	\bibfield  {author} {\bibinfo {author} {\bibfnamefont {L.~D.}\ \bibnamefont
			{Landau}}\ and\ \bibinfo {author} {\bibfnamefont {E.~M.}\ \bibnamefont
			{Lifshitz}},\ }\href {https://doi.org/10.1016/C2009-0-24487-4} {\emph
		{\bibinfo {title} {Statistical Physics}}},\ \bibinfo {edition} {3rd}\ ed.,\
	Vol.~\bibinfo {volume} {5}\ (\bibinfo  {publisher} {Elsevier Science},\
	\bibinfo {year} {1980})\BibitemShut {NoStop}%
	\bibitem [{\citenamefont {Espa{\~n}ol}(1998)}]{pE98}%
	\BibitemOpen
	\bibfield  {author} {\bibinfo {author} {\bibfnamefont {P.}~\bibnamefont
			{Espa{\~n}ol}},\ }\href {https://doi.org/10.1016/S0378-4371(97)00461-5}
	{\bibfield  {journal} {\bibinfo  {journal} {Physica A}\ }\textbf {\bibinfo
			{volume} {248}},\ \bibinfo {pages} {77} (\bibinfo {year} {1998})}\BibitemShut
	{NoStop}%
	\bibitem [{\citenamefont {Balakrishnan}\ \emph {et~al.}(2014)\citenamefont
		{Balakrishnan}, \citenamefont {Garcia}, \citenamefont {Donev},\ and\
		\citenamefont {Bell}}]{BGDB14}%
	\BibitemOpen
	\bibfield  {author} {\bibinfo {author} {\bibfnamefont {K.}~\bibnamefont
			{Balakrishnan}}, \bibinfo {author} {\bibfnamefont {A.~L.}\ \bibnamefont
			{Garcia}}, \bibinfo {author} {\bibfnamefont {A.}~\bibnamefont {Donev}},\ and\
		\bibinfo {author} {\bibfnamefont {J.~B.}\ \bibnamefont {Bell}},\ }\href
	{https://doi.org/10.1103/PhysRevE.89.013017} {\bibfield  {journal} {\bibinfo
			{journal} {Phys. Rev. E}\ }\textbf {\bibinfo {volume} {89}},\ \bibinfo
		{pages} {013017} (\bibinfo {year} {2014})}\BibitemShut {NoStop}%
	\bibitem [{\citenamefont {{\"O}ttinger}(2005)}]{hcO05}%
	\BibitemOpen
	\bibfield  {author} {\bibinfo {author} {\bibfnamefont {H.~C.}\ \bibnamefont
			{{\"O}ttinger}},\ }\href {https://doi.org/10.1002/0471727903} {\emph
		{\bibinfo {title} {Beyond Equilibrium Thermodynamics}}}\ (\bibinfo
	{publisher} {Wiley},\ \bibinfo {year} {2005})\BibitemShut {NoStop}%
	\bibitem [{\citenamefont {{\"O}ttinger}(2007)}]{hcO07}%
	\BibitemOpen
	\bibfield  {author} {\bibinfo {author} {\bibfnamefont {H.~C.}\ \bibnamefont
			{{\"O}ttinger}},\ }\href {https://doi.org/10.1557/mrs2007.191} {\bibfield
		{journal} {\bibinfo  {journal} {MRS Bull.}\ }\textbf {\bibinfo {volume}
			{32}},\ \bibinfo {pages} {936} (\bibinfo {year} {2007})}\BibitemShut
	{NoStop}%
	\bibitem [{\citenamefont {Mielke}\ \emph {et~al.}(2014)\citenamefont {Mielke},
		\citenamefont {Peletier},\ and\ \citenamefont {Renger}}]{MPR14}%
	\BibitemOpen
	\bibfield  {author} {\bibinfo {author} {\bibfnamefont {A.}~\bibnamefont
			{Mielke}}, \bibinfo {author} {\bibfnamefont {M.~A.}\ \bibnamefont
			{Peletier}},\ and\ \bibinfo {author} {\bibfnamefont {D.~R.~M.}\ \bibnamefont
			{Renger}},\ }\href {https://doi.org/10.1007/s11118-014-9418-5} {\bibfield
		{journal} {\bibinfo  {journal} {Potential Anal.}\ }\textbf {\bibinfo {volume}
			{41}},\ \bibinfo {pages} {1293} (\bibinfo {year} {2014})}\BibitemShut
	{NoStop}%
	\bibitem [{\citenamefont {Mielke}\ \emph {et~al.}(2016)\citenamefont {Mielke},
		\citenamefont {Peletier},\ and\ \citenamefont
		{Renger}}]{MielkePeletierRenger16}%
	\BibitemOpen
	\bibfield  {author} {\bibinfo {author} {\bibfnamefont {A.}~\bibnamefont
			{Mielke}}, \bibinfo {author} {\bibfnamefont {M.~A.}\ \bibnamefont
			{Peletier}},\ and\ \bibinfo {author} {\bibfnamefont {D.~R.~M.}\ \bibnamefont
			{Renger}},\ }\href@noop {} {\bibfield  {journal} {\bibinfo  {journal}
			{Journal of Non-Equilibrium Thermodynamics}\ }\textbf {\bibinfo {volume}
			{41}},\ \bibinfo {pages} {141} (\bibinfo {year} {2016})}\BibitemShut
	{NoStop}%
	\bibitem [{\citenamefont {Valsson}\ \emph {et~al.}(2016)\citenamefont
		{Valsson}, \citenamefont {Tiwary},\ and\ \citenamefont {Parrinello}}]{VTP16}%
	\BibitemOpen
	\bibfield  {author} {\bibinfo {author} {\bibfnamefont {O.}~\bibnamefont
			{Valsson}}, \bibinfo {author} {\bibfnamefont {P.}~\bibnamefont {Tiwary}},\
		and\ \bibinfo {author} {\bibfnamefont {M.}~\bibnamefont {Parrinello}},\
	}\href {https://doi.org/10.1146/annurev-physchem-040215-112229} {\bibfield
		{journal} {\bibinfo  {journal} {Annu. Rev. Phys. Chem.}\ }\textbf {\bibinfo
			{volume} {67}},\ \bibinfo {pages} {159} (\bibinfo {year} {2016})}\BibitemShut
	{NoStop}%
	\bibitem [{\citenamefont {Evans}\ and\ \citenamefont {Morriss}(2007)}]{EM07}%
	\BibitemOpen
	\bibfield  {author} {\bibinfo {author} {\bibfnamefont {D.~J.}\ \bibnamefont
			{Evans}}\ and\ \bibinfo {author} {\bibfnamefont {G.~P.}\ \bibnamefont
			{Morriss}},\ }\href {https://doi.org/10.1017/CBO9780511535307} {\emph
		{\bibinfo {title} {Statistical Mechanics of Nonequilibrium Liquids}}},\
	\bibinfo {edition} {2nd}\ ed.\ (\bibinfo  {publisher} {ANU E Press},\
	\bibinfo {year} {2007})\BibitemShut {NoStop}%
	\bibitem [{\citenamefont {Kr{\"o}ger}\ and\ \citenamefont
		{{\"O}ttinger}(2004)}]{KO04}%
	\BibitemOpen
	\bibfield  {author} {\bibinfo {author} {\bibfnamefont {M.}~\bibnamefont
			{Kr{\"o}ger}}\ and\ \bibinfo {author} {\bibfnamefont {H.~C.}\ \bibnamefont
			{{\"O}ttinger}},\ }\href {https://doi.org/10.1016/j.jnnfm.2003.11.010}
	{\bibfield  {journal} {\bibinfo  {journal} {J. Non-Newton. Fluid Mech.}\
		}\textbf {\bibinfo {volume} {120}},\ \bibinfo {pages} {175} (\bibinfo {year}
		{2004})}\BibitemShut {NoStop}%
	\bibitem [{\citenamefont {Ilg}\ \emph {et~al.}(2009)\citenamefont {Ilg},
		\citenamefont {{\"O}ttinger},\ and\ \citenamefont {Kr{\"o}ger}}]{IOK09}%
	\BibitemOpen
	\bibfield  {author} {\bibinfo {author} {\bibfnamefont {P.}~\bibnamefont
			{Ilg}}, \bibinfo {author} {\bibfnamefont {H.~C.}\ \bibnamefont
			{{\"O}ttinger}},\ and\ \bibinfo {author} {\bibfnamefont {M.}~\bibnamefont
			{Kr{\"o}ger}},\ }\href {https://doi.org/10.1103/PhysRevE.79.011802}
	{\bibfield  {journal} {\bibinfo  {journal} {Phys. Rev. E}\ }\textbf {\bibinfo
			{volume} {79}},\ \bibinfo {pages} {011802} (\bibinfo {year}
		{2009})}\BibitemShut {NoStop}%
	\bibitem [{\citenamefont {Ilg}\ \emph {et~al.}(2010)\citenamefont {Ilg},
		\citenamefont {Mavrantzas},\ and\ \citenamefont {{\"O}ttinger}}]{IMO10}%
	\BibitemOpen
	\bibfield  {author} {\bibinfo {author} {\bibfnamefont {P.}~\bibnamefont
			{Ilg}}, \bibinfo {author} {\bibfnamefont {V.}~\bibnamefont {Mavrantzas}},\
		and\ \bibinfo {author} {\bibfnamefont {H.~C.}\ \bibnamefont {{\"O}ttinger}},\
	}in\ \href {https://doi.org/10.1002/9783527630257.ch7} {\emph {\bibinfo
			{booktitle} {Modeling and Simulation in Polymers}}},\ \bibinfo {editor}
	{edited by\ \bibinfo {editor} {\bibfnamefont {P.~D.}\ \bibnamefont
			{Gujrati}}\ and\ \bibinfo {editor} {\bibfnamefont {A.~I.}\ \bibnamefont
			{Leonov}}}\ (\bibinfo  {publisher} {Wiley},\ \bibinfo {year} {2010})\
	Chap.~\bibinfo {chapter} {7}, pp.\ \bibinfo {pages} {343--383}\BibitemShut
	{NoStop}%
	\bibitem [{\citenamefont {Onsager}\ and\ \citenamefont {Machlup}(1953)}]{OM53}%
	\BibitemOpen
	\bibfield  {author} {\bibinfo {author} {\bibfnamefont {L.}~\bibnamefont
			{Onsager}}\ and\ \bibinfo {author} {\bibfnamefont {S.}~\bibnamefont
			{Machlup}},\ }\href {https://doi.org/10.1103/PhysRev.91.1505} {\bibfield
		{journal} {\bibinfo  {journal} {Phys. Rev.}\ }\textbf {\bibinfo {volume}
			{91}},\ \bibinfo {pages} {1505} (\bibinfo {year} {1953})}\BibitemShut
	{NoStop}%
	\bibitem [{\citenamefont {Eyring}(1935)}]{hE35}%
	\BibitemOpen
	\bibfield  {author} {\bibinfo {author} {\bibfnamefont {H.}~\bibnamefont
			{Eyring}},\ }\href {https://doi.org/10.1063/1.1749604} {\bibfield  {journal}
		{\bibinfo  {journal} {J. Chem. Phys.}\ }\textbf {\bibinfo {volume} {3}},\
		\bibinfo {pages} {107} (\bibinfo {year} {1935})}\BibitemShut {NoStop}%
	\bibitem [{\citenamefont {Kramers}(1940)}]{haK40}%
	\BibitemOpen
	\bibfield  {author} {\bibinfo {author} {\bibfnamefont {H.~A.}\ \bibnamefont
			{Kramers}},\ }\href {https://doi.org/10.1016/S0031-8914(40)90098-2}
	{\bibfield  {journal} {\bibinfo  {journal} {Physica}\ }\textbf {\bibinfo
			{volume} {7}},\ \bibinfo {pages} {284} (\bibinfo {year} {1940})}\BibitemShut
	{NoStop}%
	\bibitem [{\citenamefont {Bucklew}(2004)}]{jaB04}%
	\BibitemOpen
	\bibfield  {author} {\bibinfo {author} {\bibfnamefont {J.~A.}\ \bibnamefont
			{Bucklew}},\ }\href {https://doi.org/10.1007/978-1-4757-4078-3} {\emph
		{\bibinfo {title} {Introduction to Rare Event Simulation}}},\ \bibinfo
	{edition} {1st}\ ed.\ (\bibinfo  {publisher} {Springer-Verlag},\ \bibinfo
	{year} {2004})\BibitemShut {NoStop}%
	\bibitem [{\citenamefont {Vanden-Eijnden}(2006)}]{eVE06}%
	\BibitemOpen
	\bibfield  {author} {\bibinfo {author} {\bibfnamefont {E.}~\bibnamefont
			{Vanden-Eijnden}},\ }in\ \href {https://doi.org/10.1007/3-540-35273-2_13}
	{\emph {\bibinfo {booktitle} {Computer Simulations in Condensed Matter
				Systems: From Materials to Chemical Biology Volume 1}}},\ \bibinfo {editor}
	{edited by\ \bibinfo {editor} {\bibfnamefont {M.}~\bibnamefont {Ferrario}},
		\bibinfo {editor} {\bibfnamefont {G.}~\bibnamefont {Ciccotti}},\ and\
		\bibinfo {editor} {\bibfnamefont {K.}~\bibnamefont {Binder}}}\ (\bibinfo
	{publisher} {Springer},\ \bibinfo {address} {Berlin, Heidelberg},\ \bibinfo
	{year} {2006})\ pp.\ \bibinfo {pages} {453--493}\BibitemShut {NoStop}%
	\bibitem [{\citenamefont {Hartmann}\ \emph {et~al.}(2014)\citenamefont
		{Hartmann}, \citenamefont {Banisch}, \citenamefont {Sarich}, \citenamefont
		{Badowski},\ and\ \citenamefont {Sch{\"u}tte}}]{HBSBS14}%
	\BibitemOpen
	\bibfield  {author} {\bibinfo {author} {\bibfnamefont {C.}~\bibnamefont
			{Hartmann}}, \bibinfo {author} {\bibfnamefont {R.}~\bibnamefont {Banisch}},
		\bibinfo {author} {\bibfnamefont {M.}~\bibnamefont {Sarich}}, \bibinfo
		{author} {\bibfnamefont {T.}~\bibnamefont {Badowski}},\ and\ \bibinfo
		{author} {\bibfnamefont {C.}~\bibnamefont {Sch{\"u}tte}},\ }\href
	{https://doi.org/10.3390/e16010350} {\bibfield  {journal} {\bibinfo
			{journal} {Entropy}\ }\textbf {\bibinfo {volume} {16}},\ \bibinfo {pages}
		{350} (\bibinfo {year} {2014})}\BibitemShut {NoStop}%
	\bibitem [{\citenamefont {Fa{\v{c}}kovec}\ \emph {et~al.}(2015)\citenamefont
		{Fa{\v{c}}kovec}, \citenamefont {Vanden-Eijnden},\ and\ \citenamefont
		{Wales}}]{FVEW15}%
	\BibitemOpen
	\bibfield  {author} {\bibinfo {author} {\bibfnamefont {B.}~\bibnamefont
			{Fa{\v{c}}kovec}}, \bibinfo {author} {\bibfnamefont {E.}~\bibnamefont
			{Vanden-Eijnden}},\ and\ \bibinfo {author} {\bibfnamefont {D.~J.}\
			\bibnamefont {Wales}},\ }\href {https://doi.org/10.1063/1.4926940} {\bibfield
		{journal} {\bibinfo  {journal} {J. Chem. Phys.}\ }\textbf {\bibinfo {volume}
			{143}},\ \bibinfo {pages} {044119} (\bibinfo {year} {2015})}\BibitemShut
	{NoStop}%
	\bibitem [{\citenamefont {Be'er}\ and\ \citenamefont {Assaf}(2016)}]{BA16}%
	\BibitemOpen
	\bibfield  {author} {\bibinfo {author} {\bibfnamefont {S.}~\bibnamefont
			{Be'er}}\ and\ \bibinfo {author} {\bibfnamefont {M.}~\bibnamefont {Assaf}},\
	}\href {https://doi.org/10.1088/1742-5468/2016/11/113501} {\bibfield
		{journal} {\bibinfo  {journal} {J. Stat. Mech. Theory Exp.}\ }\textbf
		{\bibinfo {volume} {2016}},\ \bibinfo {pages} {113501} (\bibinfo {year}
		{2016})}\BibitemShut {NoStop}%
	\bibitem [{\citenamefont {Klimontovich}(1990)}]{ylK90}%
	\BibitemOpen
	\bibfield  {author} {\bibinfo {author} {\bibfnamefont {Y.~L.}\ \bibnamefont
			{Klimontovich}},\ }\href {https://doi.org/10.1016/0378-4371(90)90142-F}
	{\bibfield  {journal} {\bibinfo  {journal} {Physica A}\ }\textbf {\bibinfo
			{volume} {163}},\ \bibinfo {pages} {515} (\bibinfo {year}
		{1990})}\BibitemShut {NoStop}%
	\bibitem [{\citenamefont {Iacus}(2008)}]{smI08}%
	\BibitemOpen
	\bibfield  {author} {\bibinfo {author} {\bibfnamefont {S.~M.}\ \bibnamefont
			{Iacus}},\ }\href {https://doi.org/10.1007/978-0-387-75839-8} {\emph
		{\bibinfo {title} {Simulation and Inference for Stochastic Differential
				Equations}}},\ \bibinfo {edition} {1st}\ ed.\ (\bibinfo  {publisher}
	{Springer-Verlag},\ \bibinfo {year} {2008})\BibitemShut {NoStop}%
	\bibitem [{\citenamefont {Fuchs}(2013)}]{cF13}%
	\BibitemOpen
	\bibfield  {author} {\bibinfo {author} {\bibfnamefont {C.}~\bibnamefont
			{Fuchs}},\ }\href {https://doi.org/10.1007/978-3-642-25969-2} {\emph
		{\bibinfo {title} {Inference for Diffusion Processes}}},\ \bibinfo {edition}
	{1st}\ ed.\ (\bibinfo  {publisher} {Springer-Verlag},\ \bibinfo {year}
	{2013})\BibitemShut {NoStop}%
	\bibitem [{Note1()}]{Note1}%
	\BibitemOpen
	\bibinfo {note} {The presence of the divergence of $D$ is due to the
		Klimontovich interpretation for the noise.}\BibitemShut {Stop}%
	\bibitem [{\citenamefont {Dawson}(1993)}]{daD93}%
	\BibitemOpen
	\bibfield  {author} {\bibinfo {author} {\bibfnamefont {D.~A.}\ \bibnamefont
			{Dawson}},\ }in\ \href {https://doi.org/10.1007/BFb0084190} {\emph {\bibinfo
			{booktitle} {{\'E}cole d'{\'E}t{\'e} de Probabilit{\'e}s de Saint-Flour
				XXI---1991}}},\ \bibinfo {series} {Lecture Notes in Math.}, Vol.\ \bibinfo
	{volume} {1541}\ (\bibinfo  {publisher} {Springer},\ \bibinfo {address}
	{Berlin},\ \bibinfo {year} {1993})\ pp.\ \bibinfo {pages}
	{1--260}\BibitemShut {NoStop}%
	\bibitem [{\citenamefont {Embacher}\ \emph {et~al.}(2018)\citenamefont
		{Embacher}, \citenamefont {Dirr}, \citenamefont {Zimmer},\ and\ \citenamefont
		{Reina}}]{EDZR18}%
	\BibitemOpen
	\bibfield  {author} {\bibinfo {author} {\bibfnamefont {P.}~\bibnamefont
			{Embacher}}, \bibinfo {author} {\bibfnamefont {N.}~\bibnamefont {Dirr}},
		\bibinfo {author} {\bibfnamefont {J.}~\bibnamefont {Zimmer}},\ and\ \bibinfo
		{author} {\bibfnamefont {C.}~\bibnamefont {Reina}},\ }\href
	{https://doi.org/10.1098/rspa.2017.0694} {\bibfield  {journal} {\bibinfo
			{journal} {Proc. Royal Soc. A: Math. Phys. Eng. Sci.}\ }\textbf {\bibinfo
			{volume} {474}},\ \bibinfo {pages} {20170694} (\bibinfo {year}
		{2018})}\BibitemShut {NoStop}%
	\bibitem [{\citenamefont {Pavliotis}(2014)}]{Pavliotis14}%
	\BibitemOpen
	\bibfield  {author} {\bibinfo {author} {\bibfnamefont {G.~A.}\ \bibnamefont
			{Pavliotis}},\ }\href {https://doi.org/10.1007/978-1-4939-1323-7} {\emph
		{\bibinfo {title} {Stochastic processes and applications}}}\ (\bibinfo
	{publisher} {Springer},\ \bibinfo {year} {2014})\BibitemShut {NoStop}%
	\bibitem [{\citenamefont {Grmela}\ and\ \citenamefont
		{{\"O}ttinger}(1997)}]{GO97}%
	\BibitemOpen
	\bibfield  {author} {\bibinfo {author} {\bibfnamefont {M.}~\bibnamefont
			{Grmela}}\ and\ \bibinfo {author} {\bibfnamefont {H.~C.}\ \bibnamefont
			{{\"O}ttinger}},\ }\href {https://doi.org/10.1103/PhysRevE.56.6633}
	{\bibfield  {journal} {\bibinfo  {journal} {Phys. Rev. E}\ }\textbf {\bibinfo
			{volume} {56}},\ \bibinfo {pages} {6620} (\bibinfo {year}
		{1997})}\BibitemShut {NoStop}%
	\bibitem [{\citenamefont {{\"O}ttinger}\ and\ \citenamefont
		{Grmela}(1997)}]{OG97}%
	\BibitemOpen
	\bibfield  {author} {\bibinfo {author} {\bibfnamefont {H.~C.}\ \bibnamefont
			{{\"O}ttinger}}\ and\ \bibinfo {author} {\bibfnamefont {M.}~\bibnamefont
			{Grmela}},\ }\href {https://doi.org/10.1103/PhysRevE.56.6633} {\bibfield
		{journal} {\bibinfo  {journal} {Phys. Rev. E}\ }\textbf {\bibinfo {volume}
			{56}},\ \bibinfo {pages} {6633} (\bibinfo {year} {1997})}\BibitemShut
	{NoStop}%
	\bibitem [{\citenamefont {Liggett}(2005)}]{tmL05}%
	\BibitemOpen
	\bibfield  {author} {\bibinfo {author} {\bibfnamefont {T.~M.}\ \bibnamefont
			{Liggett}},\ }\href {https://doi.org/10.1007/b138374} {\emph {\bibinfo
			{title} {Interacting Particle Systems}}}\ (\bibinfo  {publisher}
	{Springer-Verlag},\ \bibinfo {year} {2005})\BibitemShut {NoStop}%
	\bibitem [{\citenamefont {Gardiner}(2009)}]{cG09}%
	\BibitemOpen
	\bibfield  {author} {\bibinfo {author} {\bibfnamefont {C.}~\bibnamefont
			{Gardiner}},\ }\href@noop {} {\emph {\bibinfo {title} {Stochastic
				Methods}}},\ \bibinfo {edition} {4th}\ ed.\ (\bibinfo  {publisher}
	{Springer-Verlag},\ \bibinfo {address} {Berlin, Heidelberg},\ \bibinfo {year}
	{2009})\BibitemShut {NoStop}%
	\bibitem [{\citenamefont {Allen}\ and\ \citenamefont {Tildesley}(2017)}]{AT17}%
	\BibitemOpen
	\bibfield  {author} {\bibinfo {author} {\bibfnamefont {M.~P.}\ \bibnamefont
			{Allen}}\ and\ \bibinfo {author} {\bibfnamefont {D.~J.}\ \bibnamefont
			{Tildesley}},\ }\href {https://doi.org/10.1093/oso/9780198803195.001.0001}
	{\emph {\bibinfo {title} {Computer Simulation of Liquids}}},\ \bibinfo
	{edition} {2nd}\ ed.\ (\bibinfo  {publisher} {Oxford University Press},\
	\bibinfo {year} {2017})\BibitemShut {NoStop}%
	\bibitem [{\citenamefont {Mondello}\ and\ \citenamefont {Grest}(1997)}]{MG97}%
	\BibitemOpen
	\bibfield  {author} {\bibinfo {author} {\bibfnamefont {M.}~\bibnamefont
			{Mondello}}\ and\ \bibinfo {author} {\bibfnamefont {G.~S.}\ \bibnamefont
			{Grest}},\ }\href {https://doi.org/10.1063/1.474002} {\bibfield  {journal}
		{\bibinfo  {journal} {J. Chem. Phys.}\ }\textbf {\bibinfo {volume} {106}},\
		\bibinfo {pages} {9327} (\bibinfo {year} {1997})}\BibitemShut {NoStop}%
	\bibitem [{\citenamefont {Guo}\ \emph {et~al.}(2003)\citenamefont {Guo},
		\citenamefont {Zhang},\ and\ \citenamefont {Zhao}}]{GZZ03}%
	\BibitemOpen
	\bibfield  {author} {\bibinfo {author} {\bibfnamefont {G.-J.}\ \bibnamefont
			{Guo}}, \bibinfo {author} {\bibfnamefont {Y.-G.}\ \bibnamefont {Zhang}},\
		and\ \bibinfo {author} {\bibfnamefont {Y.-J.}\ \bibnamefont {Zhao}},\ }\href
	{https://doi.org/10.1103/PhysRevE.67.043101} {\bibfield  {journal} {\bibinfo
			{journal} {Phys. Rev. E}\ }\textbf {\bibinfo {volume} {67}},\ \bibinfo
		{pages} {043101} (\bibinfo {year} {2003})}\BibitemShut {NoStop}%
	\bibitem [{\citenamefont {Bladt}\ and\ \citenamefont
		{S{\o}rensen}(2005)}]{BS05}%
	\BibitemOpen
	\bibfield  {author} {\bibinfo {author} {\bibfnamefont {M.}~\bibnamefont
			{Bladt}}\ and\ \bibinfo {author} {\bibfnamefont {M.}~\bibnamefont
			{S{\o}rensen}},\ }\href {https://doi.org/10.1111/j.1467-9868.2005.00508.x}
	{\bibfield  {journal} {\bibinfo  {journal} {J. Royal Stat. Soc. Ser. B}\
		}\textbf {\bibinfo {volume} {67}},\ \bibinfo {pages} {395} (\bibinfo {year}
		{2005})}\BibitemShut {NoStop}%
	\bibitem [{\citenamefont {Metzner}\ \emph {et~al.}(2007)\citenamefont
		{Metzner}, \citenamefont {Dittmer}, \citenamefont {Jahnke},\ and\
		\citenamefont {Sch{\"u}tte}}]{MDJS07}%
	\BibitemOpen
	\bibfield  {author} {\bibinfo {author} {\bibfnamefont {P.}~\bibnamefont
			{Metzner}}, \bibinfo {author} {\bibfnamefont {E.}~\bibnamefont {Dittmer}},
		\bibinfo {author} {\bibfnamefont {T.}~\bibnamefont {Jahnke}},\ and\ \bibinfo
		{author} {\bibfnamefont {C.}~\bibnamefont {Sch{\"u}tte}},\ }\href
	{https://doi.org/10.1016/j.jcp.2007.07.032} {\bibfield  {journal} {\bibinfo
			{journal} {J. Comput. Phys.}\ }\textbf {\bibinfo {volume} {227}},\ \bibinfo
		{pages} {353} (\bibinfo {year} {2007})}\BibitemShut {NoStop}%
	\bibitem [{\citenamefont {Kingman}(1962)}]{jfcK62}%
	\BibitemOpen
	\bibfield  {author} {\bibinfo {author} {\bibfnamefont {J.~F.~C.}\
			\bibnamefont {Kingman}},\ }\href {https://doi.org/10.1007/BF00531768}
	{\bibfield  {journal} {\bibinfo  {journal} {Zeitschrift f{\"u}r
				Wahrscheinlichkeitstheorie und Verwandte Gebiete}\ }\textbf {\bibinfo
			{volume} {1}},\ \bibinfo {pages} {14} (\bibinfo {year} {1962})}\BibitemShut
	{NoStop}%
	\bibitem [{\citenamefont {Bovier}\ and\ \citenamefont {den
			Hollander}(2015)}]{BdH15}%
	\BibitemOpen
	\bibfield  {author} {\bibinfo {author} {\bibfnamefont {A.}~\bibnamefont
			{Bovier}}\ and\ \bibinfo {author} {\bibfnamefont {F.}~\bibnamefont {den
				Hollander}},\ }\href {https://doi.org/10.1007/978-3-319-24777-9} {\emph
		{\bibinfo {title} {Metastability}}}\ (\bibinfo  {publisher} {Springer
		International Publishing},\ \bibinfo {year} {2015})\BibitemShut {NoStop}%
	\bibitem [{\citenamefont {Di~Ges{\`u}}\ \emph {et~al.}(2016)\citenamefont
		{Di~Ges{\`u}}, \citenamefont {Leli{\`e}vre}, \citenamefont {Le~Peutrec},\
		and\ \citenamefont {Nectoux}}]{DLLN16}%
	\BibitemOpen
	\bibfield  {author} {\bibinfo {author} {\bibfnamefont {G.}~\bibnamefont
			{Di~Ges{\`u}}}, \bibinfo {author} {\bibfnamefont {T.}~\bibnamefont
			{Leli{\`e}vre}}, \bibinfo {author} {\bibfnamefont {D.}~\bibnamefont
			{Le~Peutrec}},\ and\ \bibinfo {author} {\bibfnamefont {B.}~\bibnamefont
			{Nectoux}},\ }\href {https://doi.org/10.1039/C6FD00120C} {\bibfield
		{journal} {\bibinfo  {journal} {Faraday Discuss.}\ }\textbf {\bibinfo
			{volume} {195}},\ \bibinfo {pages} {469} (\bibinfo {year}
		{2016})}\BibitemShut {NoStop}%
	\bibitem [{\citenamefont {Feng}\ and\ \citenamefont {Kurtz}(2006)}]{FK06}%
	\BibitemOpen
	\bibfield  {author} {\bibinfo {author} {\bibfnamefont {J.}~\bibnamefont
			{Feng}}\ and\ \bibinfo {author} {\bibfnamefont {T.~G.}\ \bibnamefont
			{Kurtz}},\ }\href@noop {} {\emph {\bibinfo {title} {Large Deviations for
				Stochastic Processes}}}\ (\bibinfo  {publisher} {AMS},\ \bibinfo {year}
	{2006})\BibitemShut {NoStop}%
	\bibitem [{\citenamefont {M{\"u}ller}(1994)}]{pM94}%
	\BibitemOpen
	\bibfield  {author} {\bibinfo {author} {\bibfnamefont {P.}~\bibnamefont
			{M{\"u}ller}},\ }\href {https://doi.org/10.1351/pac199466051077} {\emph
		{\bibinfo {title} {Glossary of terms used in physical organic chemistry}}},\
	Vol.~\bibinfo {volume} {66}\ (\bibinfo  {publisher} {Walter de Gruyter},\
	\bibinfo {year} {1994})\ pp.\ \bibinfo {pages} {1077--1184}\BibitemShut
	{NoStop}%
	\bibitem [{Note2()}]{Note2}%
	\BibitemOpen
	\bibinfo {note} {Also known as \protect \emph {law of mass action} \cite
		{sS87,mG12} or \protect \emph {Guldberg-Waage dynamics} \cite
		{GW867}}\BibitemShut {NoStop}%
	\bibitem [{\citenamefont {Laio}\ and\ \citenamefont {Parrinello}(2002)}]{LP02}%
	\BibitemOpen
	\bibfield  {author} {\bibinfo {author} {\bibfnamefont {A.}~\bibnamefont
			{Laio}}\ and\ \bibinfo {author} {\bibfnamefont {M.}~\bibnamefont
			{Parrinello}},\ }\href {https://doi.org/10.1073/pnas.202427399} {\bibfield
		{journal} {\bibinfo  {journal} {Proc. Natl. Acad. Sci. U.S.A.}\ }\textbf
		{\bibinfo {volume} {99}},\ \bibinfo {pages} {12562} (\bibinfo {year}
		{2002})}\BibitemShut {NoStop}%
	\bibitem [{\citenamefont {Zhang}\ \emph {et~al.}(2016)\citenamefont {Zhang},
		\citenamefont {Hartmann},\ and\ \citenamefont {Sch{\"u}tte}}]{ZHS16}%
	\BibitemOpen
	\bibfield  {author} {\bibinfo {author} {\bibfnamefont {W.}~\bibnamefont
			{Zhang}}, \bibinfo {author} {\bibfnamefont {C.}~\bibnamefont {Hartmann}},\
		and\ \bibinfo {author} {\bibfnamefont {C.}~\bibnamefont {Sch{\"u}tte}},\
	}\href {https://doi.org/10.1039/C6FD00147E} {\bibfield  {journal} {\bibinfo
			{journal} {Faraday Discuss.}\ }\textbf {\bibinfo {volume} {195}},\ \bibinfo
		{pages} {365} (\bibinfo {year} {2016})}\BibitemShut {NoStop}%
	\bibitem [{\citenamefont {Arnrich}\ \emph {et~al.}(2012)\citenamefont
		{Arnrich}, \citenamefont {Mielke}, \citenamefont {Peletier}, \citenamefont
		{Savar{\'e}},\ and\ \citenamefont {Veneroni}}]{AMPSV12}%
	\BibitemOpen
	\bibfield  {author} {\bibinfo {author} {\bibfnamefont {S.}~\bibnamefont
			{Arnrich}}, \bibinfo {author} {\bibfnamefont {A.}~\bibnamefont {Mielke}},
		\bibinfo {author} {\bibfnamefont {M.~A.}\ \bibnamefont {Peletier}}, \bibinfo
		{author} {\bibfnamefont {G.}~\bibnamefont {Savar{\'e}}},\ and\ \bibinfo
		{author} {\bibfnamefont {M.}~\bibnamefont {Veneroni}},\ }\href
	{https://doi.org/10.1007/s00526-011-0440-9} {\bibfield  {journal} {\bibinfo
			{journal} {Calc. Var. Partial Differ. Equ.}\ }\textbf {\bibinfo {volume}
			{44}},\ \bibinfo {pages} {419} (\bibinfo {year} {2012})}\BibitemShut
	{NoStop}%
	\bibitem [{\citenamefont {Joshi}(2015)}]{bJ15}%
	\BibitemOpen
	\bibfield  {author} {\bibinfo {author} {\bibfnamefont {B.}~\bibnamefont
			{Joshi}},\ }\href {https://doi.org/10.3934/dcdsb.2015.20.1077} {\bibfield
		{journal} {\bibinfo  {journal} {Discrete Continuous Dyn. Syst. Ser. B}\
		}\textbf {\bibinfo {volume} {20}},\ \bibinfo {pages} {1077} (\bibinfo {year}
		{2015})}\BibitemShut {NoStop}%
	\bibitem [{\citenamefont {Gillespie}(2001)}]{dtG01}%
	\BibitemOpen
	\bibfield  {author} {\bibinfo {author} {\bibfnamefont {D.~T.}\ \bibnamefont
			{Gillespie}},\ }\href {https://doi.org/10.1063/1.1378322} {\bibfield
		{journal} {\bibinfo  {journal} {J. Chem. Phys.}\ }\textbf {\bibinfo {volume}
			{115}},\ \bibinfo {pages} {1716} (\bibinfo {year} {2001})}\BibitemShut
	{NoStop}%
	\bibitem [{\citenamefont {Grmela}(1993)}]{mG93}%
	\BibitemOpen
	\bibfield  {author} {\bibinfo {author} {\bibfnamefont {M.}~\bibnamefont
			{Grmela}},\ }\href {https://doi.org/10.1103/PhysRevE.48.919} {\bibfield
		{journal} {\bibinfo  {journal} {Phys. Rev. E}\ }\textbf {\bibinfo {volume}
			{48}},\ \bibinfo {pages} {919} (\bibinfo {year} {1993})}\BibitemShut
	{NoStop}%
	\bibitem [{\citenamefont {Grmela}(2010)}]{mG10}%
	\BibitemOpen
	\bibfield  {author} {\bibinfo {author} {\bibfnamefont {M.}~\bibnamefont
			{Grmela}},\ }in\ \href {https://doi.org/10.1016/S0065-2377(10)39002-8} {\emph
		{\bibinfo {booktitle} {Advances in Chemical Engineering}}},\ \bibinfo
	{series} {Advances in Chemical Engineering}, Vol.~\bibinfo {volume} {39},\
	\bibinfo {editor} {edited by\ \bibinfo {editor} {\bibfnamefont {D.~H.}\
			\bibnamefont {West}}\ and\ \bibinfo {editor} {\bibfnamefont {G.}~\bibnamefont
			{Yablonsky}}}\ (\bibinfo  {publisher} {Academic Press},\ \bibinfo {year}
	{2010})\ Chap.~\bibinfo {chapter} {2}, pp.\ \bibinfo {pages}
	{75--129}\BibitemShut {NoStop}%
	\bibitem [{\citenamefont {Montefusco}(2019)}]{aM19}%
	\BibitemOpen
	\bibfield  {author} {\bibinfo {author} {\bibfnamefont {A.}~\bibnamefont
			{Montefusco}},\ }\emph {\bibinfo {title} {Dynamic Coarse-Graining via
			Large-Deviation Theory}},\ \href {https://doi.org/10.3929/ethz-b-000354751}
	{Ph.D. thesis},\ \bibinfo  {school} {ETH Z{\"u}rich} (\bibinfo {year}
	{2019})\BibitemShut {NoStop}%
	\bibitem [{\citenamefont {Mielke}(2011)}]{aM11}%
	\BibitemOpen
	\bibfield  {author} {\bibinfo {author} {\bibfnamefont {A.}~\bibnamefont
			{Mielke}},\ }\href {https://doi.org/10.1088/0951-7715/24/4/016} {\bibfield
		{journal} {\bibinfo  {journal} {Nonlinearity}\ }\textbf {\bibinfo {volume}
			{24}},\ \bibinfo {pages} {1329} (\bibinfo {year} {2011})}\BibitemShut
	{NoStop}%
	\bibitem [{\citenamefont {{\"O}ttinger}(2015)}]{hcO15}%
	\BibitemOpen
	\bibfield  {author} {\bibinfo {author} {\bibfnamefont {H.~C.}\ \bibnamefont
			{{\"O}ttinger}},\ }\href {https://doi.org/10.1103/PhysRevE.91.032147}
	{\bibfield  {journal} {\bibinfo  {journal} {Phys. Rev. E}\ }\textbf {\bibinfo
			{volume} {91}},\ \bibinfo {pages} {032147} (\bibinfo {year}
		{2015})}\BibitemShut {NoStop}%
	\bibitem [{\citenamefont {Gillespie}(2000)}]{dtG00}%
	\BibitemOpen
	\bibfield  {author} {\bibinfo {author} {\bibfnamefont {D.~T.}\ \bibnamefont
			{Gillespie}},\ }\href {https://doi.org/10.1063/1.481811} {\bibfield
		{journal} {\bibinfo  {journal} {J. Chem. Phys.}\ }\textbf {\bibinfo {volume}
			{113}},\ \bibinfo {pages} {297} (\bibinfo {year} {2000})}\BibitemShut
	{NoStop}%
	\bibitem [{Note3()}]{Note3}%
	\BibitemOpen
	\bibinfo {note} {The diffusion approximation may be found by three heuristic
		arguments: the first corresponds to expanding the generator to first order
		in~$1/n$, the second to considering the Kramers-Moyal expansion in the
		equation for the law, and the third to replacing the Poisson noise, which
		describes the jumps, by a Brownian one for large~$n$ \cite
		{ngvK83,EK05,AK11}.}\BibitemShut {Stop}%
	\bibitem [{\citenamefont {Bhattacharjee}\ \emph {et~al.}(2015)\citenamefont
		{Bhattacharjee}, \citenamefont {Balakrishnan}, \citenamefont {Garcia},
		\citenamefont {Bell},\ and\ \citenamefont {Donev}}]{BBGBD15}%
	\BibitemOpen
	\bibfield  {author} {\bibinfo {author} {\bibfnamefont {A.~K.}\ \bibnamefont
			{Bhattacharjee}}, \bibinfo {author} {\bibfnamefont {K.}~\bibnamefont
			{Balakrishnan}}, \bibinfo {author} {\bibfnamefont {A.~L.}\ \bibnamefont
			{Garcia}}, \bibinfo {author} {\bibfnamefont {J.~B.}\ \bibnamefont {Bell}},\
		and\ \bibinfo {author} {\bibfnamefont {A.}~\bibnamefont {Donev}},\ }\href
	{https://doi.org/10.1063/1.4922308} {\bibfield  {journal} {\bibinfo
			{journal} {J. Chem. Phys.}\ }\textbf {\bibinfo {volume} {142}},\ \bibinfo
		{pages} {224107} (\bibinfo {year} {2015})}\BibitemShut {NoStop}%
	\bibitem [{Note4()}]{Note4}%
	\BibitemOpen
	\bibinfo {note} {An estimation of the equilibration time~$t_2$ should be done
		separately, for instance with the help of the various methods available in
		the literature \cite {DLLN16,afV98,BLS15}. In our case, the ratio $t_2/\Delta
		t$ is of the order of $20$ \cite [Figure~6.4]{aM19}.}\BibitemShut {Stop}%
	\bibitem [{Note5()}]{Note5}%
	\BibitemOpen
	\bibinfo {note} {We choose a small value $\xi ^+_1$. The positive values $\xi
		^+_j$ are in geometric progression starting from $\xi ^+_1$, and the negative
		ones are $\xi ^-_j = - \xi ^+_j$.}\BibitemShut {Stop}%
	\bibitem [{\citenamefont {Mirrahimi}\ and\ \citenamefont
		{Souganidis}(2013)}]{MS13}%
	\BibitemOpen
	\bibfield  {author} {\bibinfo {author} {\bibfnamefont {S.}~\bibnamefont
			{Mirrahimi}}\ and\ \bibinfo {author} {\bibfnamefont {P.~E.}\ \bibnamefont
			{Souganidis}},\ }\href {https://doi.org/10.1007/s00030-012-0156-3} {\bibfield
		{journal} {\bibinfo  {journal} {Nonlinear Differ. Equat. Appl.}\ }\textbf
		{\bibinfo {volume} {20}},\ \bibinfo {pages} {129} (\bibinfo {year}
		{2013})}\BibitemShut {NoStop}%
	\bibitem [{\citenamefont {Bruna}\ and\ \citenamefont {Chapman}(2012)}]{BC12}%
	\BibitemOpen
	\bibfield  {author} {\bibinfo {author} {\bibfnamefont {M.}~\bibnamefont
			{Bruna}}\ and\ \bibinfo {author} {\bibfnamefont {S.~J.}\ \bibnamefont
			{Chapman}},\ }\href {https://doi.org/10.1063/1.4767058} {\bibfield  {journal}
		{\bibinfo  {journal} {J. Chem. Phys.}\ }\textbf {\bibinfo {volume} {137}},\
		\bibinfo {pages} {204116} (\bibinfo {year} {2012})}\BibitemShut {NoStop}%
	\bibitem [{\citenamefont {Touchette}(2011)}]{hT11}%
	\BibitemOpen
	\bibfield  {author} {\bibinfo {author} {\bibfnamefont {H.}~\bibnamefont
			{Touchette}},\ }\href@noop {} {\bibfield  {journal} {\bibinfo  {journal}
			{ArXiv e-prints}\ } (\bibinfo {year} {2011})},\ \Eprint
	{https://arxiv.org/abs/1106.4146} {arXiv:1106.4146 [cond-mat.stat-mech]}
	\BibitemShut {NoStop}%
	\bibitem [{\citenamefont {Sieniutycz}(1987)}]{sS87}%
	\BibitemOpen
	\bibfield  {author} {\bibinfo {author} {\bibfnamefont {S.}~\bibnamefont
			{Sieniutycz}},\ }\href {https://doi.org/10.1016/0009-2509(87)87020-3}
	{\bibfield  {journal} {\bibinfo  {journal} {Chem. Eng. Sci.}\ }\textbf
		{\bibinfo {volume} {42}},\ \bibinfo {pages} {2697} (\bibinfo {year}
		{1987})}\BibitemShut {NoStop}%
	\bibitem [{\citenamefont {Grmela}(2012)}]{mG12}%
	\BibitemOpen
	\bibfield  {author} {\bibinfo {author} {\bibfnamefont {M.}~\bibnamefont
			{Grmela}},\ }\href {https://doi.org/10.1016/j.physd.2012.02.008} {\bibfield
		{journal} {\bibinfo  {journal} {Physica D}\ }\textbf {\bibinfo {volume}
			{241}},\ \bibinfo {pages} {976} (\bibinfo {year} {2012})}\BibitemShut
	{NoStop}%
	\bibitem [{\citenamefont {Waage}\ and\ \citenamefont {Guldberg}(1867)}]{GW867}%
	\BibitemOpen
	\bibfield  {author} {\bibinfo {author} {\bibfnamefont {P.}~\bibnamefont
			{Waage}}\ and\ \bibinfo {author} {\bibfnamefont {C.~M.}\ \bibnamefont
			{Guldberg}},\ }\href {http://www.sudoc.fr/098378155} {\emph {\bibinfo {title}
			{{\'E}tudes sur les affinit{\'e}s chimiques}}}\ (\bibinfo  {publisher}
	{Br{\o}gger et Christie},\ \bibinfo {address} {Christiania},\ \bibinfo {year}
	{1867})\ p.~\bibinfo {pages} {74}\BibitemShut {NoStop}%
	\bibitem [{\citenamefont {van Kampen}(1983)}]{ngvK83}%
	\BibitemOpen
	\bibfield  {author} {\bibinfo {author} {\bibfnamefont {N.~G.}\ \bibnamefont
			{van Kampen}},\ }in\ \href@noop {} {\emph {\bibinfo {booktitle}
			{Thermodynamics and kinetics of biological processes}}},\ \bibinfo {editor}
	{edited by\ \bibinfo {editor} {\bibfnamefont {I.}~\bibnamefont {Lamprecht}}\
		and\ \bibinfo {editor} {\bibfnamefont {A.~I.}\ \bibnamefont {Zotin}}}\
	(\bibinfo  {publisher} {Walter de Gruyter},\ \bibinfo {year} {1983})\ pp.\
	\bibinfo {pages} {181--195}\BibitemShut {NoStop}%
	\bibitem [{\citenamefont {Ethier}\ and\ \citenamefont {Kurtz}(2005)}]{EK05}%
	\BibitemOpen
	\bibfield  {author} {\bibinfo {author} {\bibfnamefont {S.~N.}\ \bibnamefont
			{Ethier}}\ and\ \bibinfo {author} {\bibfnamefont {T.~G.}\ \bibnamefont
			{Kurtz}},\ }\href {https://doi.org/10.1002/9780470316658} {\emph {\bibinfo
			{title} {Markov Processes: Characterization and Convergence}}},\ \bibinfo
	{edition} {2nd}\ ed.\ (\bibinfo  {publisher} {John Wiley \& Sons},\ \bibinfo
	{year} {2005})\BibitemShut {NoStop}%
	\bibitem [{\citenamefont {Anderson}\ and\ \citenamefont {Kurtz}(2011)}]{AK11}%
	\BibitemOpen
	\bibfield  {author} {\bibinfo {author} {\bibfnamefont {D.~F.}\ \bibnamefont
			{Anderson}}\ and\ \bibinfo {author} {\bibfnamefont {T.~G.}\ \bibnamefont
			{Kurtz}},\ }in\ \href {https://doi.org/10.1007/978-1-4419-6766-4} {\emph
		{\bibinfo {booktitle} {Design and Analysis of Biomolecular Circuits}}},\
	\bibinfo {editor} {edited by\ \bibinfo {editor} {\bibfnamefont
			{H.}~\bibnamefont {Koeppl}}, \bibinfo {editor} {\bibfnamefont
			{D.}~\bibnamefont {Densmore}}, \bibinfo {editor} {\bibfnamefont
			{G.}~\bibnamefont {Setti}},\ and\ \bibinfo {editor} {\bibfnamefont
			{M.}~\bibnamefont {di~Bernardo}}}\ (\bibinfo  {publisher} {Springer},\
	\bibinfo {year} {2011})\ Chap.~\bibinfo {chapter} {1}, pp.\ \bibinfo {pages}
	{3--42}\BibitemShut {NoStop}%
	\bibitem [{\citenamefont {Voter}(1998)}]{afV98}%
	\BibitemOpen
	\bibfield  {author} {\bibinfo {author} {\bibfnamefont {A.~F.}\ \bibnamefont
			{Voter}},\ }\href {https://doi.org/10.1103/PhysRevB.57.R13985} {\bibfield
		{journal} {\bibinfo  {journal} {Phys. Rev. B}\ }\textbf {\bibinfo {volume}
			{57}},\ \bibinfo {pages} {R13985} (\bibinfo {year} {1998})}\BibitemShut
	{NoStop}%
	\bibitem [{\citenamefont {Binder}\ \emph {et~al.}(2015)\citenamefont {Binder},
		\citenamefont {Leli{\`e}vre},\ and\ \citenamefont {Simpson}}]{BLS15}%
	\BibitemOpen
	\bibfield  {author} {\bibinfo {author} {\bibfnamefont {A.}~\bibnamefont
			{Binder}}, \bibinfo {author} {\bibfnamefont {T.}~\bibnamefont
			{Leli{\`e}vre}},\ and\ \bibinfo {author} {\bibfnamefont {G.}~\bibnamefont
			{Simpson}},\ }\href {https://doi.org/10.1016/j.jcp.2015.01.002} {\bibfield
		{journal} {\bibinfo  {journal} {J. Comput. Phys.}\ }\textbf {\bibinfo
			{volume} {284}},\ \bibinfo {pages} {595} (\bibinfo {year}
		{2015})}\BibitemShut {NoStop}%
\end{thebibliography}

\end{document}